# Applications of Information Theory in Plant Disease Management[1]

Gareth Hughes

School of Biological Sciences, University of Edinburgh, Edinburgh EH9 3JG, UK

*We seek the truth, but, in this life, we usually have to be content with probability distributions.*
William A. Benish

## 1. Introduction

This section is intended to set the scene for the applications of information theory discussed in sections 2, 3 and 4. This is an abbreviated version of part of Chapter 11 of Madden et al. (2007), which can be consulted by those who seek a more detailed treatment of the introductory material.

### 1.1. Disease management

The first step in the formulation of a disease management strategy for any particular cropping system is to identify the most important risk factors among those on the long list of possible candidates. This is facilitated by basic epidemiological studies of pathogen life cycles, and an understanding of the way in which weather and cropping factors affect the level of initial inoculum and the rate of progress through the pathogen life cycle. In order to be able to identify the most important risk factors, we need information both on the candidate risk factors and on the definitive status of the crops in which they are studied. In the widely-used epidemiological terminology for such analyses, a crop that definitively required treatment is referred to as a *case*, one that definitively did not is referred to as a *control*. Note that the classification into cases and controls must be independent of risk factors that might be used as a basis for decision-making. In crop protection, this is often a retrospective classification of untreated crops based on yield or an end-of-season disease assessment.

Once the important risk factors have been identified, we need a way of combining them so that we can use data on risk factors to make a prediction of whether or not crop protection measures are required. We use the term *risk algorithm* to refer to a calculation that combines data on identified risk factors in order to make an assessment of the need for crop protection measures. In plant disease management, a risk algorithm characterizes the relationship between a binary response variable (i.e., the requirement for crop protection measures or otherwise) and one or more risk factors (i.e., the explanatory variables). Typically, a risk algorithm is formulated from a data set where the requirement for crop protection measures, or otherwise, has been assessed retrospectively (and so definitively) for a number of crops for which various risk factors thought likely to be related to disease intensity and/or crop yield reduction have also been measured.

The disease management decision-making problem can now be outlined as follows. We wish to classify crops as requiring protection measures or otherwise. Because the objective is to use the crop protection measures to prevent disease developing to an economically significant level, we cannot measure this requirement directly. Instead, we predict the requirement, using data relating to the important risk factors, combined via a risk algorithm. The risk algorithm generates an *indicator variable* (or, for brevity, simply an indicator) that provides a basis for a *diagnostic test* (or, for brevity, simply a test; *predictor* is synonymous) relating to the need for crop protection measures. The diagnostic test comprises an indicator variable and a classification rule. A classification rule is a threshold (or a set of thresholds) on the scale of the indicator variable.

## 1.2. Example 1: potato late blight

Johnson et al. (1996, 1998) analyzed relationships between weather factors and epidemics of potato late blight (caused by *Phytophthora infestans*) in south-central Washington, USA. Their objective was to develop a forecasting model to aid in disease management at the regional scale. Using retrospective data from the study area over the period 1970 – 1994, Johnson et al. (1996) identified important risk factors for potato late blight in south-central Washington as follows: whether or not the previous year was an outbreak year; the number of days on which rain fell during April and May; and the total precipitation during May when the daily minimum temperature was ≤5°C. Any year in the study period in which late blight was confirmed in any field in the study area was classified as an outbreak year (these are the cases), otherwise a year was classified as a non-outbreak year (these are the controls). The procedures outlined for distinguishing between outbreak years and non-outbreak years are assumed to have been free of error.

Johnson et al. (1996) derived indicator variables by combining the data for risk factors using both discriminant function analysis and logistic regression analysis. Johnson et al. (1996) described diagnostic tests based on their discriminant function analysis and on their logistic regression analysis that resulted in the same classification of years as either outbreak or non-outbreak. Both tests correctly identified 11/12 outbreak years (cases) and 11/13 non-outbreak years (controls). Results of this kind are often shown in a two-way table, as follows.

| | True status | |
|---|---|---|
| | Outbreak year (cases) | Non-outbreak year (controls) |
| Predicted status | | |
| Outbreak year | 11 | 2 |
| Non-outbreak year | 1 | 11 |
| Total | 12 | 13 |

In this example, whichever of the two methods of formulating a diagnostic test is used, most outbreak years are correctly classified (these are true positives) and so are most non-outbreak years (these are true negatives). However, a small number of classification errors arise. Outbreak years wrongly classified as non-outbreak years are false negatives, while non-outbreak years wrongly classified as outbreak years are false positives. The true positive proportion (*TPP*) expresses the number of true positives as a proportion of the total number of cases. The true negative proportion (*TNP*) expresses the number of true negatives as a proportion of the total number of controls. The false negative proportion is $FNP = 1 - TPP$, and the false positive proportion is $FPP = 1 - TNP$.

*TPP* is often referred to as the *sensitivity* and *TNP* as the *specificity* of a test. For the potato late blight data set, sensitivity = 11/12 = 0.92 and specificity = 11/13 = 0.85. Sensitivity and specificity represent two kinds of accuracy, respectively, for cases and controls. Sensitivity and specificity are independent of the proportions of cases and controls in a data set and can therefore be viewed as properties of a test.

### 1.3. Example 2: Sclerotinia stem rot of oil seed rape

Researchers at the Swedish University of Agricultural Sciences analyzed relationships between weather, crop and disease factors and epidemics of Sclerotinia stem rot of spring sown oil seed rape (caused by *Scelrotinia sclerotiorum*) in Sweden (Yuen et al., 1996; Twengström et al., 1998). Their objective was to develop a forecasting model for predicting the need for fungicide applications, working at the field scale. Stepwise logistic regression procedures were used to identify risk factors for Sclerotinia stem rot of spring sown oil seed rape. Twengström et al. (1998) identified six important risk factors for Sclerotinia stem rot of spring sown oil seed rape in Sweden, as follows: number of years since the previous oilseed rape crop; percentage disease incidence in the previous host crop; plant population density; rainfall in the previous two weeks; weather forecast; and regional risk for apothecium development.

Data from about 800 fields, untreated with fungicide, were collected over a ten year period and used in the analysis. Retrospectively, the fields were divided into two groups: those with >25% diseased plants (these are the cases) and those with ≤25% diseased plants (these are the controls). Because a sampling procedure was used to assess % diseased plants, there is uncertainty attached to the classification of cases and controls. We will, however, assume that cases and controls have been classified definitively and correctly.

Risk algorithms for Sclerotinia stem rot were based on logistic regression analysis (Yuen et al., 1996; Twengström et al., 1998). A simplified risk points scale was derived from the estimated regression coefficients in order to facilitate practical application of the risk algorithm (Twengström et al., 1998). The risk points accruing to each individual crop were accumulated to provide an overall risk score, and then frequency distributions of risk scores were plotted separately for crops with >25% diseased plants (cases) and crops with ≤25% diseased plants (controls). In this example, there is overlap between the two distributions in the range 25 – 60 risk points (see Twengström et al., 1998). This means that for the Sclerotinia stem rot data it is more difficult than for the potato late blight data in Example 1 (section 1.2) to choose an appropriate classification rule. The choice of a threshold value on the risk points scale is not so straightforward because the extent of the overlap between frequency distributions for cases and controls means that there is no unequivocal best choice of threshold for distinguishing between crops with >25% diseased plants and crops with ≤25% diseased plants. For example, the outcomes of choosing threshold risk scores of 35, 40 and 50 points are as follows.

| Threshold risk score | *TPP* (sensitivity) | *TNP* (specificity) | *FNP* | *FPP* |
|---|---|---|---|---|
| 35 | 0.89 | 0.77 | 0.11 | 0.23 |
| 40 | 0.76 | 0.84 | 0.24 | 0.14 |
| 50 | 0.34 | 0.95 | 0.66 | 0.05 |

If a relatively low threshold risk score is adopted, this yields a test with higher sensitivity (*TPP* is increased, *FNP* decreased) and a lower specificity (*TNP* is decreased, *FPP* increased). Conversely, adopting a relatively high threshold risk score yields a test with higher specificity (*TNP* is increased, *FPP* decreased) and a lower sensitivity (*TPP* is decreased, *FNP* increased). Note that there is no way of altering the threshold such that both *FNP* and *FPP* are simultaneously decreased.

## 1.4.    Sensitivity and specificity

Sensitivity (*TPP*) and specificity (*TNP*) are characteristics of a diagnostic test. They can be written in the form of conditional probabilities, as follows.

- *TPP* is an estimate of the conditional probability $\Pr(T_1|D_1)$ (read as 'the probability of an indicator score above the threshold value, given that the true status of the crop was a case (i.e., requiring treatment)').

- *TNP* is an estimate of the conditional probability $\Pr(T_2|D_2)$ (read as 'the probability of an indicator score at or below the threshold value, given that the true status of the crop was a control (i.e., not requiring treatment)').

- *FNP* is an estimate of the conditional probability $\Pr(T_2|D_1)$.

- *FPP* is an estimate of the conditional probability $\Pr(T_1|D_2)$.

An alternative formulation of these characteristics is provided by the calculation of two *likelihood ratios*, as follows (Go, 1998). The likelihood ratio of an indicator score above the threshold value is defined here as:

$$LR_1 = \frac{\Pr(T_1|D_1)}{\Pr(T_1|D_2)} = \frac{TPP}{1-TNP} = \frac{TPP}{FPP}.$$

The likelihood ratio of an indicator score below the threshold value is:

$$LR_2 = \frac{\Pr(T_2|D_1)}{\Pr(T_2|D_2)} = \frac{1-TPP}{TNP} = \frac{FNP}{TNP}.$$

For a perfect test (*TPP* = *TNP* = 1), $LR_1$ is indefinitely large and $LR_2$ is zero (Go, 1998). For a test with no discriminatory power, both $LR_1$ and $LR_2$ are equal to one (Go, 1998). $LR_1 > 1$ and $LR_2 < 1$ are the minimum requirements of a useful test. Ideally, we would like to have $LR_1$ as large as possible and simultaneously have $LR_2$ as small as possible (Biggerstaff, 2000).

For the potato late blight indicator described in Example 1, the test with *TPP* = 0.92 and *TNP* = 0.85 has the corresponding likelihood ratios $LR_1$ = 6.13 and $LR_2$ = 0.09. For the Sclerotinia stem rot indicator described in Example 2, the test with *TPP* = 0.89 and *TNP* = 0.77 has $LR_1$ = 3.87 and $LR_2$ = 0.14. Biggerstaff (2000) gives a useful geometrical interpretation of $LR_1$ and $LR_2$ and their relationship to the sensitivity and specificity of a diagnostic test.

## 1.5.    Prior and posterior probabilities

The discussion so far has been not been concerned directly with the problem of predicting whether or not a crop requires treatment. When sensitivity and specificity are written as conditional probabilities (section 1.4), it can be seen that the conditionality is such that the probability of the test result is given, conditional on the true disease status (case or control) of the crop in question. When we are forecasting plant disease, we do not know the true disease status of the crop in question. Assuming we have developed an appropriate test, and applied it, what we know is the test result ($T_1$ or $T_2$) and what we wish to do is calculate the probability of a crop's requirement for treatment (or otherwise), conditional on this test result. In order to accomplish this, we need to know the characteristics of the diagnostic test (section 1.4) and the unconditional (pre-test) probability of a crop's requirement for treatment, denoted $\Pr(D_1)$, based in some appropriate way on our previous experience (note that the pre-test probability of no requirement for treatment $\Pr(D_2) = 1 - \Pr(D_1)$). $\Pr(D_1)$ and $\Pr(D_2)$ are referred to as *prior probabilities*. *Bayes' theorem* facilitates the updating of prior probabilities, using evidence from risk factors, to *posterior probabilities*.

In practice, the calculations are simplified if we work in terms of odds rather than probabilities, where $odds(\text{event}) = \dfrac{\Pr(\text{event})}{1 - \Pr(\text{event})}$ . Then, from Bayes' theorem:

$$odds(D_1|T_1) = odds(D_1) \cdot LR_1$$

from which we can see that as long as $LR_1 > 1$, the effect of a test that provides a prediction of need for treatment is to increase the posterior odds of need for treatment, relative to the prior odds. The posterior odds $odds(D_1|T_1)$ can be converted to a posterior probability using $\Pr(\text{event}) = \dfrac{odds(\text{event})}{1 + odds(\text{event})}$ . The posterior probability $\Pr(D_1|T_1)$ is referred to as the *positive predictive value* (*PPV*). Also from Bayes' theorem:

$$odds(D_2|T_2) = odds(D_2)/LR_2$$

from which we can see that as long as $LR_2 < 1$, the effect of a test that provides a prediction that treatment will not be required is to increase the posterior odds that treatment will not be required, relative to the prior odds. It is now clear why $LR_1 > 1$ and $LR_2 < 1$ are the minimum requirements of a useful indicator (section 1.4). The posterior odds $odds(D_2|T_2)$ can be converted to the posterior probability $\Pr(D_2|T_2)$, which is referred to as the *negative predictive value* (*NPV*). *PPV* and *NPV* are not inherent properties of a diagnostic test, since they depend on the corresponding prior probabilities.

Other useful statements of Bayes' theorem as applied in the context of disease management are available. The posterior odds of requirement for treatment following a prediction that treatment will not be required is:

$$odds(D_1|T_2) = odds(D_1) \cdot LR_2$$

which, when converted to a probability, is equal to $1 - NPV$. The posterior odds that treatment will not be required following a prediction that treatment will be required is:

$$odds(D_2|T_1) = odds(D_2)/LR_1$$

which is equal to $1 - PPV$ when converted to a probability.

Continuing with Example 1 (see section 1.2), outbreaks of potato late blight were identified in 12 of 25 years in commercial potato fields in south-central Washington between 1970 and 1994 (Johnson et al., 1996,). About half the years in the study period were outbreak years. We therefore take $\Pr(D_1) = 0.5$ to be a reasonable estimate of the prior probability of an outbreak. We have previously calculated the sensitivity (= 0.92) and specificity (= 0.85) for a diagnostic test (section 1.2). Now, using Bayes' theorem to move from the prior odds of an outbreak year to the posterior odds of an outbreak year, given a prediction of an outbreak year, we have:

$$odds(D_1|T_1) = \frac{0.5}{1-0.5} \cdot \frac{0.92}{1-0.85} = 6.13 \text{ and then, if required,}$$

$$\Pr(D_1|T_1) = \frac{6.13}{1+6.13} = 0.86 .$$

Before using the test, the evidence suggests that outbreaks are about as likely to occur as not to occur. However, if the test results in a prediction of an outbreak year, probability of an outbreak is increased to 0.86.

Similarly, to move from the prior odds of a non-outbreak year to the posterior odds of a non-outbreak year, given a prediction of a non-outbreak year:

$$odds\left(D_2\middle|T_2\right) = \frac{0.5}{1-0.5} \cdot \left(\frac{1-0.92}{0.85}\right)^{-1} = 10.63 \text{ and then, if required,}$$

$$\Pr\left(D_2\middle|T_2\right) = \frac{10.63}{1+10.63} = 0.91 \,.$$

Before using the test, it appears that outbreaks are about as likely to occur as not to occur but, if the test results in a prediction of a non-outbreak year, the probability of a non-outbreak year is increased to 0.91.

Continuing with Example 2 (see section 1.3), a 20-year average for the frequency of need for control measures for Sclerotinia stem rot of oil seed rape in Uppland, east-central Sweden, is 16% (Yuen and Hughes, 2002). We therefore take $\Pr(D+) = 0.16$ as the prior probability of need for fungicide application (so the prior odds is $odds(D_1) = 0.19$). Different threshold values of the risk score give different values of sensitivity and specificity (section 1.3). These data can now be combined, via Bayes' theorem, to characterize predictive values, as follows.

| Threshold risk score | Need for treatment | | | |
| --- | --- | --- | --- | --- |
| | $LR_1$ | Prior odds | Posterior odds | *PPV* |
| 35 | 3.87 | 0.19 | 0.74 | 0.42 |
| 40 | 4.75 | 0.19 | 0.90 | 0.48 |
| 50 | 6.80 | 0.19 | 1.30 | 0.56 |

| Threshold risk score | No need for treatment | | | |
| --- | --- | --- | --- | --- |
| | $LR_2$ | Prior odds | Posterior odds | *NPV* |
| 35 | 0.14 | 5.25 | 36.75 | 0.97 |
| 40 | 0.29 | 5.25 | 18.38 | 0.95 |
| 50 | 0.69 | 5.25 | 7.56 | 0.88 |

## 1.6. Rare events

While the advantages of being able to predict the occurrence of particularly rare events, or the non-occurrence of particularly common ones, hardly need spelling out, we can see from Bayes' theorem that in practice it is difficult to make useful predictions where rare events are concerned. For a prior probability that is very low, or very high, it is difficult to obtain posterior probabilities that will change a decision maker's view (based on this prior probability), even if a test with good sensitivity and specificity characteristics is available. Thus diseases that occur very frequently or very infrequently pose a problem from the point of view of prediction (see section 2.3).

## 2.    Information theory

Information theory, insofar as it concerns us here, is a branch of probability and statistics involving the analysis of communications. A prediction (forecast) that is the output of a risk algorithm (section 1.1) is a kind of communication, generically referred to as a *message*. Information theory enables us to analyze and quantify the information content of messages such as predictions made in the context of plant disease management and related disciplines. What follows is not intended as a general introduction to information theory. Cover and Thomas (2006, chapter 1) and Theil (1967, chapters 1-3) are useful in that context, and the latter provides the basis for the following sections 2.1 and 2.2.

### 2.1.    The information content of a message

Consider an event $E$ that occurs with probability $\Pr(E)$, $0 \leq \Pr(E) \leq 1$. If we receive a *definite message* (i.e., a message that is, without question, correct) that $E$ has occurred, the information content of this message is:

$$h[\Pr(E)] = \log\left[\frac{1}{\Pr(E)}\right] \qquad (1)$$

(Theil, 1967). Thus, the information content of a definite message is a function only of the probability of occurrence of event $E$ before the message is received (i.e., the prior probability). The function $h[\Pr(E)]$ is continuous, and decreases monotonically from $\infty$ to 0 as $\Pr(E)$ increases from 0 to 1. In the latter case, if before receiving the message we are already certain that $E$ will occur, then the information content of the message that $E$ has occurred is zero. In the former case, there is an indefinitely large amount of information in any message which assigns a positive probability to an event whose prior probability was zero (Hobson and Cheng, 1973). The choice of base of logarithm does not matter other than to define the units of information. If logarithms base 2 are chosen, the unit of information is the bit. For natural logarithms and logarithms base 10, the units of information are, respectively, the nit and the Hartley. Given that L. P. Hartley's great novel of information transmission *The Go-Between* was published in 1953, just a few years after Claude Shannon and Warren Weaver's *The Mathematical Theory of Communication*, one does rather wish that a unit of information had been named in his honour. More prosaically, but no less deservedly, it is the early work of R. V. L. Hartley on the mathematics of information transmission that is recognized.

The prior probability of an outbreak of potato late blight in commercial fields in south-central Washington was taken to be $\Pr(D_1)=0.5$ (section 1.5). Working in natural logarithms, the information content of a message that an outbreak had occurred is then $\ln(1/0.5)=0.69$ nits. Note that working in logarithms base 2, the information content in this case is equal to 1 bit, or in logarithms base 10, 0.3 Hartleys. For Sclerotinia stem rot of oil seed rape in east-central Sweden, $\Pr(D_1)$ was taken to be 0.16 (section 1.5). In this case (working in natural logarithms), the information content of a message that an outbreak had occurred is $\ln(1/0.16)=1.83$ nits. The information content of this message is higher because the prior probability of disease is lower than in the case of the potato late blight example. The circumstances and opinions of a person receiving a message such as this one are not part of the calculation of information content. Thus, the message has the same information content whether it is received by you, me, or a farmer in Uppsala with a field of oil seed rape to cultivate. Note also that the information calculations in this chapter do not include factors related to the costs and benefits of forecasts made in the context of plant disease management, nor the relative values of the different kinds of errors that may occur when a forecast is made. As Somoza and Mossman (1992) have noted, epidemiological applications of information theory tend to emphasize the role of diagnostic testing in reducing uncertainty rather than determining choice of treatment.

Not all the messages we receive will be definite messages. Perhaps more often, we will receive an *indefinite message*, one that serves to transform a set of prior probabilities into a corresponding set of posterior probabilities. In this case, we can generalize Equation 1 as follows:

information content of message

$$= \log \left[ \frac{\text{probability of the event after the message is received}}{\text{probability of the event before the message is received}} \right] \qquad (2)$$

(Theil, 1967). Using Equation 2, we can calculate the information content of the forecasts of need for treatment of Sclerotinia stem rot of oil seed rape (section 1.3), for three different threshold risk scores. From Equation 2, we have:

information content of message $T_1 = h\big[\Pr(D_1)\big] - h\big[\Pr(D_1 \mid T_1)\big]$,

values of which are given in Table 1. We can regard the information content calculated in this way (sometimes called *information gain*) as a measure of the value of the forecast. It is positive when the posterior probability exceeds the prior probability, zero if the two probabilities are equal, and negative if the posterior probability is smaller than the prior. Obviously it is the first of these three outcomes that we require from a useful forecast of the need for treatment.

Similarly, we can calculate the information content of the forecasts of no need for treatment of Sclerotinia stem rot, for three different threshold risk scores. From Equation 2, we now have:

information content of message $T_2 = h\big[\Pr(D_2)\big] - h\big[\Pr(D_2 \mid T_2)\big]$,

values of which are given in Table 2. Again the information content calculated in this way is positive when the posterior probability exceeds the prior probability, zero if the two probabilities are equal and negative if the posterior probability is smaller than the prior.

**Table 1.** Sclerotinia stem rot of oil seed rape. Using data from section 1.5, values of the information content (nits) of a message that transforms the prior probability $\Pr(D_1)$ to the posterior probability $\Pr(D_1|T_1)$ for risk score thresholds of 35, 40 and 50 points are calculated here.

| Threshold risk score | Need for treatment | | |
| --- | --- | --- | --- |
| | Prior probability | Posterior probability | Information content (nits) |
| 35 | 0.16 | 0.42 | 0.98 |
| 40 | 0.16 | 0.48 | 1.08 |
| 50 | 0.16 | 0.56 | 1.25 |

**Table 2.** Sclerotinia stem rot of oil seed rape. Using data from section 1.5, values of the information content (nits) of a message that transforms the prior probability $\Pr(D_2)$ to the posterior probability $\Pr(D_2|T_2)$ for risk score thresholds of 35, 40 and 50 points are calculated here.

| Threshold risk score | No need for treatment | | |
| | Prior probability | Posterior probability | Information content (nits) |
| --- | --- | --- | --- |
| 35 | 0.84 | 0.97 | 0.14 |
| 40 | 0.84 | 0.95 | 0.12 |
| 50 | 0.84 | 0.88 | 0.05 |

The calculations of information content as shown in Tables 1 and 2 illustrate two points. First, that use of a threshold of 50 points results in the largest information gain from the message $T_1$, while use of a threshold of 35 points results in the largest information gain from the message $T_2$. Second, it is apparent that the information gains from the message $T_1$ are rather larger than the information gains from the message $T_2$, whatever the choice of threshold. This is because information gain is a measure of the value of a forecast given what is already known, and in this case $\Pr(D_2)$, the prior probability of no need for treatment, is considerably larger than $\Pr(D_1)$, the prior probability of need for treatment . For any forecast, it is useful to know the probability that it will be correct, and reassuring if this probability is reasonably close to 1. But the fact that a forecast of no need for treatment provides 97% correct decisions (with a threshold risk score of 35) needs to be set against the fact that a policy of taking no notice of the forecast and never treating would lead to 84% correct decisions anyway (Table 2). A forecast of need for treatment that provides 56% correct decisions (with a threshold risk score of 50) may be some way short of perfection, but it is a considerable improvement on the 16% correct decisions to treat that would be made on the basis of ignoring the forecast and always treating (Table 1). In the case of forecasts of need for treatment, the correct forecast percentage is smaller than in the case of forecasts of no need for treatment, but the information gain is larger.

## 2.2. Expected information content

In Examples 1 and 2 (sections 1.2 and 1.3), the true status of a crop may be either $D_1$ (denoting a disease outbreak, or the need for treatment) or $D_2$ (denoting no outbreak, or no need for treatment). Here, we adopt an extended notation, in which the true status of a crop may take any one of $m$ states, $D_1 \ldots D_j \ldots D_m$. The corresponding probabilities are $\Pr(D_1) \ldots \Pr(D_j) \ldots \Pr(D_m)$, and:

$$\sum_{j=1}^{m} \Pr(D_j) = 1$$

$$\Pr(D_j) \geq 0 .$$

When we receive a definite message that a crop has true status $D_j$, the information content of this message (from Equation 1) is:

$$h\big[\Pr(D_j)\big] = \log\left[\frac{1}{\Pr(D_j)}\right] .$$

We cannot calculate this quantity until the message is received, because the message "$D_j$ occurred" may refer to any one of $D_1 \ldots D_j \ldots D_m$. We can, however, calculate the *expected information content* before the message is received. This is the weighted average of the $h[\text{Pr}(D_j)]$ values. Since the message "$D_j$ occurred" is received with probability $\text{Pr}(D_j)$, the expected information content (denoted $H[\text{Pr}(D)]$) is:

$$H\big[\text{Pr}(D)\big] = \sum_{j=1}^{m} \text{Pr}\big(D_j\big) \log\left[\frac{1}{\text{Pr}(D_j)}\right] \tag{3}$$

The expected information content of a probability distribution – in this case, the distribution of $D$ – is often referred to as the *entropy* of that distribution and written:

$$H\big[\text{Pr}(D)\big] = -\sum_{j=1}^{m} \text{Pr}\big(D_j\big) \log\big[\text{Pr}(D_j)\big].$$

We note that $H[\text{Pr}(D)] \geq 0$ and take $\text{Pr}(D_j)\log[\text{Pr}(D_j)] = 0$ if $\text{Pr}(D_j)$=0. If any $\text{Pr}(D_j) = 1$, $H[\text{Pr}(D)] = 0$. This is reasonable since we expect nothing from the forecast if we are already certain of the outcome. $H[\text{Pr}(D)]$ has its maximum value when all the $\text{Pr}(D_j)$ have the same value, equal to $1/n$ (Theil, 1967). This is also reasonable, since a message that tells us what actually happened when all outcomes have the same prior probability will have a larger information content than when some outcomes have larger prior probabilities than others. For example, for forecasts of potato late blight outbreaks in commercial fields in south-central Washington (section 1.2), the probability of an outbreak is $\text{Pr}(D_1)$=0.5 and the probability of no outbreak is $\text{Pr}(D_2)$=0.5. Working in natural logarithms, the expected information content of a message that tells us what happened is (from Equation 3) $H[\text{Pr}(D)] = 0.5\ln(2) + 0.5\ln(2) = 0.69$ nits. For forecasts of Sclerotinia stem rot of oil seed rape in east-central Sweden (section 1.3), the probability of need for treatment $\text{Pr}(D_1)$=0.16 and the probability of no need for treatment $\text{Pr}(D_2)$=0.84. In this case, the expected information content of a message that tells us what happened is (from Equation 3) $H[\text{Pr}(D)] = 0.16\ln(6.25) + 0.84\ln(1.19) = 0.44$ nits. The greater the uncertainty prevailing before a message is received, the larger is the expected information content of a message that tells us what happened.

We need now to generalize Equation 3 to be able to calculate expected information content for an indefinite message. Recall that the true status of a crop may be any of $D_1 \ldots D_j \ldots D_m$. The corresponding probabilities are $\text{Pr}(D_1) \ldots \text{Pr}(D_j) \ldots \text{Pr}(D_m)$. A message (denoted $T$) is received which serves to transform these prior probabilities into the posterior probabilities $\text{Pr}(D_1|T) \ldots \text{Pr}(D_j|T) \ldots \text{Pr}(D_m|T)$, where:

$$\sum_{j=1}^{m} \text{Pr}\big(D_j|T\big) = 1$$

$$\text{Pr}\big(D_j|T\big) \geq 0.$$

When we receive the message $T$, the information content of this message (from Equation 2) is:

$$\text{information content of message } T = \log\left[\frac{\text{Pr}(D_j|T)}{\text{Pr}(D_j)}\right].$$

The expected information content of the message $T$ (denoted $I[\text{Pr}(D_j|T):\text{Pr}(D_j)]$) is the weighted average of the information contents, the weights being the posterior probabilities $\text{Pr}(D_j|T)$:

$$I\big[\Pr(D_j|T):\Pr(D_j)\big] = \sum_{j=1}^{m}\Pr(D_j|T)\log\left[\frac{\Pr(D_j|T)}{\Pr(D_j)}\right] \tag{4}$$

(Theil, 1967). The quantity $I[\Pr(D_j|T):\Pr(D_j)]) \geq 0$, and is equal to zero if and only if $\Pr(D_j|T) = \Pr(D_j)$, $j = 1\ldots m$. Thus the expected information content of a message which leaves the prior probabilities unchanged is zero, which is reasonable.

In the terminology of Kullback (1968), the quantity $I[\Pr(D_j|T):\Pr(D_j)]$) is a *directed divergence*. In the terminology of Cover and Thomas (2006) $I[\Pr(D_j|T):\Pr(D_j)]$) is the *relative entropy* between the posterior and prior probability distributions (sometimes, though not by Kullback, colloquially referred to as the *Kullback-Leibler distance*). Its application as a measure of diagnostic information has been discussed by Benish (1999)[2].

As an illustration, we use the predictor from Example 2 (section 1.3), with a threshold risk points score of 50. The required posterior probabilities were given in section 1.5, and are repeated here in Table 3. Information contents in nits are calculated using Equation 2, with $\Pr(D_1) = 0.16$ and $\Pr(D_2) = 0.84$, and expected information contents in nits then calculated from Equation 4 (see Table 3). For this implementation of the predictor, the expected information content of prediction $T_1$ (need for treatment) is much larger than that of prediction $T_2$ (no need for treatment) (Table 3, see also Fig. 1C).

One way of interpreting the expected information from a particular prediction $T_i$ is to note that the quantity $I[\Pr(D_j|T_i):\Pr(D_j)]$) can be written as:

$$I\big[\Pr(D_j|T_i):\Pr(D_j)\big] = \sum_{j=1}^{m}\Pr(D_j|T_i)\log\left[\frac{1}{\Pr(D_j)}\right] - \sum_{j=1}^{m}\Pr(D_j|T_i)\log\left[\frac{1}{\Pr(D_j|T_i)}\right]$$

which shows that $I[\Pr(D_j|T_i):\Pr(D_j)]$) is a calculation of the reduction in entropy of the distribution of $D$ attributable to the prediction $T_i$ (at a given prior probability) (see Benish, 1999). Numerically, this yields $I[\Pr(D_j|T_1):\Pr(D_j)]) = 1.110 - 0.685 = 0.425$ nits for prediction $T_1$ in the current example (with $\Pr(D_1) = 0.16$). For prediction $T_2$ in the current example, $I[\Pr(D_j|T_2):\Pr(D_j)]) = 0.368 - 0.360 = 0.008$ nits (also with $\Pr(D_1) = 0.16$).

## 2.3. Benish's information graphs

Here, notwithstanding our more flexible notation, we restrict our attention to the situation as in Examples 1 and 2. Thus the true status of a crop, $D_j$ ($j = 1\ldots m$) is described in one of two categories (so $m=2$). $D_1$ denotes that the true status is a disease outbreak, or the need for treatment. $D_2$ denotes that the true status is no outbreak, or no need for treatment. The predicted status, $T_i$ ($i = 1\ldots n$) is also described in one of two categories (so $n=2$), based on an indicator variable that is the output of a risk algorithm. An appropriate threshold risk score is adopted and values of the indicator variable above this threshold score are taken as forecasts of a disease outbreak, or the need for treatment (denoted here $T_1$). Values of the indicator variable at or below this threshold score are taken as forecasts of no outbreak, or no need for treatment (denoted here $T_2$).

---

[2] Benish also explains clearly why the difference between the entropy of the distribution of prior probabilities and the entropy of the distribution of posterior probabilities is *not* a good measure of diagnostic information (see, for example, Mossman and Somoza (1992)).

**Table 3.** Data for the Sclerotinia stem rot predictor with threshold risk score = 50, with prior probability Pr($D_1$)=0.16. A. Posterior probabilities (from section 1.5). B. Information contents.

A.

| Prediction, $T_i$ | Posterior probability | | Row sums |
|---|---|---|---|
| | Pr($D_1|T_i$) | Pr($D_2|T_i$) | |
| $T_1$ | 0.564 | 0.436 | 1 |
| $T_2$ | 0.117 | 0.883 | 1 |

B.

| Prediction, $T_i$ | Information content (nits) | | Expected information (nits) |
|---|---|---|---|
| | Ln[Pr($D_1|T_i$)/ Pr($D_1$)] | Ln[Pr($D_2|T_i$)/ Pr($D_2$)] | $I$[Pr($D_j|T_i$):Pr($D_j$)] |
| $T_1$ | 1.260 | -0.656 | 0.425 |
| $T_2$ | -0.314 | 0.050 | 0.008 |

A *prediction-realization table* (Theil, 1967) has the probabilities Pr($D_j$) and Pr($T_i$) in the margins of the table. In the body of the table are the joint probabilities Pr($T_i \cap D_j$) (Table 4). Generally, Pr($T_i \cap D_j$) = Pr($T_i|D_j$)Pr($D_j$) = Pr($D_j|T_i$)Pr($T_i$) (which amounts to a statement of Bayes' theorem). However, a special case arises if the occurrence of event $D_j$ does not affect the probability of event $T_i$. In this case, Pr($T_i|D_j$) = Pr($T_i$), and $T_i$ and $D_j$ are independent events. Then Pr($T_i \cap D_j$) = Pr($T_i$)Pr($D_j$). This special case applies to the undesirable situation in which a prediction is of no value in determining the true status.

**Table 4.** The prediction-realization table for a predictor with two categories of true status $D_j$, $j$=1..$m$, $m$=2, and two categories of predicted status, $T_i$, $i$=1..$n$, $n$=2. In the body of the table are the joint probabilities Pr($T_i \cap D_j$).

| Prediction, $T_i$ | Realization, $D_j$ | | Row sums |
|---|---|---|---|
| | $D_1$ | $D_2$ | |
| $T_1$ | Pr($T_1 \cap D_1$) | Pr($T_1 \cap D_2$) | Pr($T_1$) |
| $T_2$ | Pr($T_2 \cap D_1$) | Pr($T_2 \cap D_2$) | Pr($T_2$) |
| Column sums | Pr($D_1$) | Pr($D_2$) | 1 |

In the general case, posterior probabilities Pr($D_j|T_i$) are calculated as follows:

$$\Pr\left(D_j|T_i\right) = \frac{\Pr\left(T_i|D_j\right)\Pr\left(D_j\right)}{\Pr\left(T_i\right)} \tag{5}$$

in which, in practice, Pr($T_i$) is found via the Law of Total Probability from Pr($T_i$) = Pr($T_i|D_1$)Pr($D_1$) + Pr($T_i|D_2$)Pr($D_2$) (Table 4).

Working in natural logarithms, and abbreviating $I$[Pr($D_j|T_1$):Pr($D_j$)]) to $I$($T_1$), the expected information content in nits for prediction $T_1$ is, from Equation 4:

$$I(T_1) = \sum_{j=1}^{m} \Pr(D_j|T_1)\ln\left[\frac{\Pr(D_j|T_1)}{\Pr(D_j)}\right]$$

$$= \Pr(D_1|T_1)\ln\left[\frac{\Pr(D_1|T_1)}{\Pr(D_1)}\right] + \Pr(D_2|T_1)\ln\left[\frac{\Pr(D_2|T_1)}{\Pr(D_2)}\right]$$

(6)

Then:

$$I(T_1) = \frac{\Pr(D_1)\Pr(T_1|D_1)}{\Pr(T_1)}\left[\ln\left(\frac{\Pr(T_1|D_1)}{\Pr(T_1)}\right)\right] + \frac{\Pr(D_2)\Pr(T_1|D_2)}{\Pr(T_1)}\left[\ln\left(\frac{\Pr(T_1|D_2)}{\Pr(T_1)}\right)\right]$$

and after some rearrangement:

$$I(T_1) = \frac{\Pr(D_1)\Pr(T_1|D_1)}{\Pr(T_1)}\left[\ln(\Pr(T_1|D_1))\right] + \frac{\Pr(D_2)\Pr(T_1|D_2)}{\Pr(T_1)}\left[\ln(\Pr(T_1|D_2))\right] - \ln(\Pr(T_1))$$

(7)

which is equation 5 from Benish (2002) (but note that Benish works in logarithms base 2). Thus for $m=2$, $I(T_1)$ is a function of the prior probability $\Pr(D_1)$ (since $\Pr(D_2) = 1-\Pr(D_1)$) and the conditional probabilities $\Pr(T_1|D_1)$ and $\Pr(T_1|D_2)$, which we recognize as, respectively, the true positive proportion (*TPP*, sensitivity) and false positive proportion (*FPP*, 1−specificity) of the predictor (section 1.4). Note that in the case where prediction $T_1$ is a definite message, $\Pr(T_1|D_1) = 1$, $\Pr(T_1|D_2) = 0$ (and we take $0\ln(0) = 0$), and Equation 7 reduces to:

$$I(T_1) = \ln\left[\frac{1}{\Pr(D_1)}\right] = h\left[\Pr(D_1)\right]$$

which we recognize as the information content of $T_1$ in nits (Equation 1).

Now, similarly, abbreviating $I[\Pr(D_j|T_2):\Pr(D_j)]$ to $I(T_2)$, we have:

$$I(T_2) = \sum_{j=1}^{m} \Pr(D_j|T_2)\ln\left[\frac{\Pr(D_j|T_2)}{\Pr(D_j)}\right]$$

$$= \Pr(D_1|T_2)\ln\left[\frac{\Pr(D_1|T_2)}{\Pr(D_1)}\right] + \Pr(D_2|T_2)\ln\left[\frac{\Pr(D_2|T_2)}{\Pr(D_2)}\right]$$

(8)

and it can then be shown that:

$$I(T_2) = \frac{\Pr(D_1)\Pr(T_2|D_1)}{\Pr(T_2)}\left[\ln(\Pr(T_2|D_1))\right] + \frac{\Pr(D_2)\Pr(T_2|D_2)}{\Pr(T_2)}\left[\ln(\Pr(T_2|D_2))\right] - \ln(\Pr(T_2)).$$

Thus for $m=2$, $I(T_2)$ is a function of the prior probability $\Pr(D_1)$ and the conditional probabilities $\Pr(T_2|D_1)$ and $\Pr(T_2|D_2)$, which we recognize as, respectively, the false negative proportion (*FNP*, 1−sensitivity) and the true negative proportion (*TNP*, specificity) of the predictor (section 1.4). Noting that $\Pr(T_2) = 1-\Pr(T_1)$), the expected information content for prediction $T_2$ in nits can be written:

$$I(T_2) = \frac{\Pr(D_1)\Pr(T_2|D_1)}{1-\Pr(T_1)}\left[\ln(\Pr(T_2|D_1))\right] + \frac{\Pr(D_2)\Pr(T_2|D_2)}{1-\Pr(T_1)}\left[\ln(\Pr(T_2|D_2))\right] - \ln(1-\Pr(T_1))$$

(9)

which is equation 6 from Benish (2002) (again note that Benish elects to work in logarithms base 2). In the case where prediction $T_2$ is a definite message, $\Pr(T_2|D_1) = 0$, $\Pr(T_2|D_2) = 1$ (and we take $0\ln(0) = 0$), and Equation 9 reduces to:

$$I(T_2) = \ln\left[\frac{1}{\Pr(D_2)}\right] = h[\Pr(D_2)]$$

which is the information content of $T_2$ in nits (Equation 1).

Equations 7 and 9 can be used to plot graphs showing relationships between the expected information contents $I(T_1)$ and $I(T_2)$ and the prior probability (in the present notation) $\Pr(D_1)$ (Benish, 2002). Fig. 1 shows such graphs for the predictor of Sclerotinia stem rot of oil seed rape described in Example 2. Expected information contents are zero when $\Pr(D_1) = 0$ or $\Pr(D_1) = 1$, since we are already certain of the outcome. Intermediate values of $\Pr(D_1)$ at which the expected information contents $I(T_1)$ and $I(T_2)$ are at their respective maximum values can be found by differentiating Equations 7 and 9, equating the resulting expressions to zero, and solving for $\Pr(D_1)$. The maximum value of $I(T_1)$ occurs at the value of $\Pr(D_1)$ given by:

$$\Pr(D_1) = \frac{FPP\left[TPP\left(\ln\left(\dfrac{TPP}{FPP}\right) - 1\right) + FPP\right]}{J^2} \tag{10}$$

and the maximum value of $I(T_2)$ occurs at the value of $\Pr(D_1)$ given by:

$$\Pr(D_1) = \frac{TNP\left[FNP\left(\ln\left(\dfrac{FNP}{TNP}\right) - 1\right) + TNP\right]}{J^2} \tag{11}$$

where $J$ is Youden's index, $J = TPP+TNP-1$ (Youden, 1950). The corresponding (maximum) values of $I(T_1)$ and $I(T_2)$ can be found by substitution back into Equations 7 and 9, respectively.

From Fig. 1 we can see that when a threshold risk score of 50 is adopted, the expected information content of a prediction of need for treatment, given that the prior probability of need for treatment is $\Pr(D_1) = 0.16$, is $I(T_1) = 0.425$ nits, and that this is not far from the maximum value of $I(T_1) = 0.438$ nits. Generally, the expected information content of a prediction of need for treatment at $\Pr(D_1) = 0.16$ is larger than the expected information content of a prediction of no need for treatment, both in absolute terms and relative to the maximum expected information contents of the respective predictions.

Fig. 1 characterizes different versions of the predictor of Sclerotinia stem rot of oil seed rape described in Example 2, but in a rather abstract way. It would be helpful to be able to take the further step of translating the expected information content of a prediction, given a prior probability, into a posterior probability. To achieve this, first recall Equations 6 and 8, respectively:

$$I(T_1) = \Pr(D_1|T_1)\ln\left[\frac{\Pr(D_1|T_1)}{\Pr(D_1)}\right] + \Pr(D_2|T_1)\ln\left[\frac{\Pr(D_2|T_1)}{\Pr(D_2)}\right]$$

and

$$I(T_2) = \Pr(D_1|T_2)\ln\left[\frac{\Pr(D_1|T_2)}{\Pr(D_1)}\right] + \Pr(D_2|T_2)\ln\left[\frac{\Pr(D_2|T_2)}{\Pr(D_2)}\right].$$

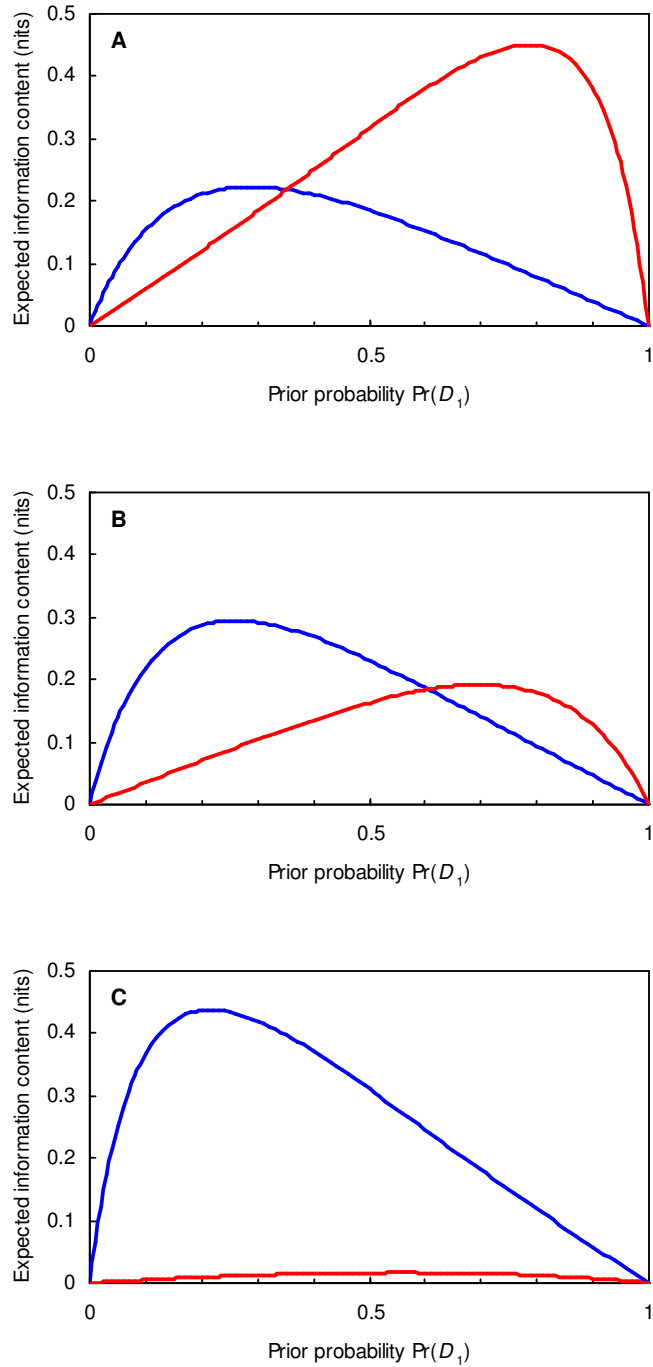

**FIG. 1.** Information graphs for the predictor of Sclerotinia stem rot of oil seed rape described in Example 2, $I(T_1)$ in blue, $I(T_2)$ in red. **A.** Threshold risk score = 35. The maximum value of $I(T_1) = 0.223$ nits occurs at $\Pr(D_1) = 0.287$. The maximum value of $I(T_2) = 0.45$ nits occurs at $\Pr(D_1) = 0.788$. At $\Pr(D_1) = 0.16$, $I(T_1) = 0.196$ nits and $I(T_2) = 0.096$ nits. **B.** Threshold risk score = 40. The maximum value of $I(T_1) = 0.294$ nits occurs at $\Pr(D_1) = 0.26$. The maximum value of $I(T_2) = 0.192$ nits occurs at $\Pr(D_1) = 0.698$. At $\Pr(D_1) = 0.16$, $I(T_1) = 0.27$ nits and $I(T_2) = 0.057$ nits. **C.** Threshold risk score = 50. The maximum value of $I(T_1) = 0.438$ nits occurs at $\Pr(D_1) = 0.215$. The maximum value of $I(T_2) = 0.017$ nits occurs at $\Pr(D_1) = 0.56$. At $\Pr(D_1) = 0.16$, $I(T_1) = 0.425$ nits and $I(T_2) = 0.008$ nits.

Values are tabulated by Kullback (1968). Now, if we adopt a particular value of the prior probability $\Pr(D_1)$ (noting that $\Pr(D_1) + \Pr(D_2) = 1$), we have a relationships between the expected information content of an indirect message and posterior probability (noting that $\Pr(D_1|T_1) + \Pr(D_2|T_1) = 1$ and $\Pr(D_1|T_2) + \Pr(D_2|T_2) = 1$). Fig. 2 shows the relationship between $\Pr(D_1|T_1)$ and $I(T_1)$ for $\Pr(D_1) = 0.16$, as in Example 2. From Fig. 2 it can be seen that if the expected information is equal to zero, the posterior probability is equal to the prior probability, which is reasonable. The posterior probability $\Pr(D_1|T_1)$ corresponding to $I(T_1)$ is found from the intersection of the vertical line $I(T_1) = 0.27$ nits (see Fig. 1B) with the part of the curve that is above the prior probability $\Pr(D_1) = 0.16$. This gives $\Pr(D_1|T_1) = 0.48$, and so $\Pr(D_2|T_1) = 0.52$ by subtraction. The posterior probability $\Pr(D_1|T_2)$ corresponding to $I(T_2)$ is found from the intersection of the vertical line $I(T_2) = 0.057$ nits (see Fig. 1B) with the part of the curve that is below the prior probability $\Pr(D_1) = 0.16$. This gives $\Pr(D_1|T_2) = 0.05$, and so $\Pr(D_2|T_2) = 0.95$ by subtraction.

In passing, note that we can look at the problem of predicting rare events (section 1.6) in terms of expected information. First, consider a hypothetical predictor with high sensitivity ($TPP = 0.9$) and specificity ($TNP = 0.9$) and suppose that a review of the available evidence in relation to the occurrence of a disease leads to the conclusion that the prior probability of need for fungicide application is $\Pr(D_1) = 0.05$. After a prediction of need for fungicide application, the posterior probability of this need is $\Pr(D_1|T_1) = 0.32$ (from Bayes' theorem). The posterior probability of need for a fungicide application exceeds the prior probability by about a factor of six (0.32 compared with 0.05) following a prediction of this need. However, the end result is that about two-thirds of the predictions of a requirement for fungicide application will be for crops that do not actually require it, since $\Pr(D_2|T_1) = 1 - \Pr(D_1|T_1) = 0.68$. Now view this problem in terms of expected information. Fig. 3 shows the relationship between $\Pr(D_1|T_1)$ and $I(T_1)$ for $\Pr(D_1) = 0.05$. The expected information content of $T_1$, a prediction of need for fungicide application, is 0.37 nits (from Equation 6), and the resulting posterior probability of need for fungicide application $\Pr(D_1|T_1) = 0.32$ (from Fig. 3). The value of $I(T_1)$ ($\approx 1$ nit) required to raise the posterior probability of need for fungicide application above $\Pr(D_1|T_1) = 0.5$ is large when considered in the context of currently available predictors used in plant disease management.

## 2.4. Mutual information

We continue, for the moment, to restrict our attention to the situation in which the true status of a crop, $D_j$ ($j = 1…m$), is described in one of two categories (so $m=2$). $D_1$ denotes that the true status is a disease outbreak, or the need for treatment. $D_2$ denotes that the true status is no outbreak, or no need for treatment. The predicted status, $T_i$ ($i = 1…n$), is also described in one of two categories (so $n=2$). $T_1$ denotes a prediction of a disease outbreak, or the need for treatment. $T_2$ denotes a prediction of no outbreak, or no need for treatment. Recalling Table 4, we define the *mutual information $h_{ij}$* as:

$$h_{ij} = \log\left[\frac{\Pr(T_i \cap D_j)}{\Pr(T_i)\Pr(D_j)}\right] \tag{12}$$

If $T_i$ and $D_j$ are independent events, $h_{ij} = 0$. Otherwise,

$$h_{ij} = \log\left[\frac{\Pr(D_j|T_i)\Pr(T_i)}{\Pr(T_i)\Pr(D_j)}\right] = \log\left[\frac{\Pr(D_j|T_i)}{\Pr(D_j)}\right] \tag{13}$$

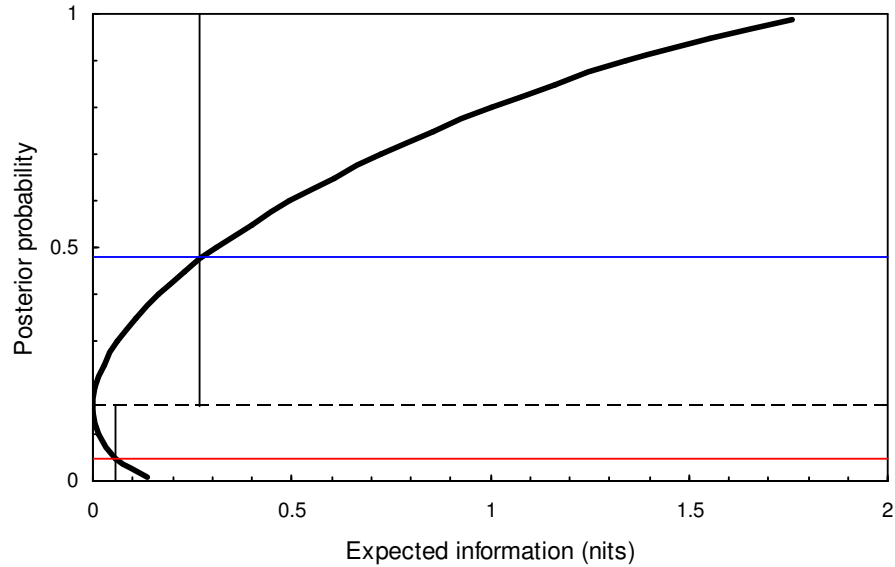

**FIG. 2.** The relationship between posterior probability and the expected information content of an indirect message. This illustration is based on the information graph for the predictor of Sclerotinia stem rot of oil seed rape described in Fig. 1B. A horizontal dashed line represents the prior probability $\Pr(D_1) = 0.16$, at which $I(T_1) = 0.27$ nits and $I(T_2) = 0.057$ nits (see Fig. 1B). The vertical line from $I(T_1) = 0.27$ nits intersects the curve above $\Pr(D_1) = 0.16$ at $\Pr(D_1|T_1) = 0.48$ (as indicated by the horizontal blue line). The vertical line from $I(T_2) = 0.057$ nits intersects the curve below $\Pr(D_1) = 0.16$ at $\Pr(D_1|T_2) = 0.05$ (as indicated by the horizontal red line). Then $\Pr(D_2|T_1) = 1 - \Pr(D_1|T_1) = 0.52$ and $\Pr(D_2|T_2) = 1 - \Pr(D_1|T_2) = 0.95$ (see also section 1.5).

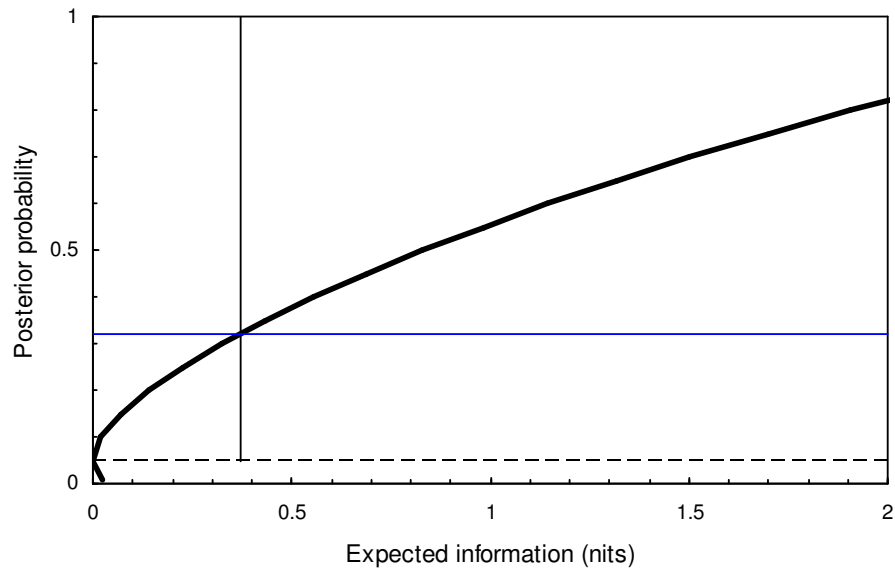

**FIG. 3.** The relationship between posterior probability and the expected information content of an indirect message. This illustration is based on a hypothetical predictor with high sensitivity ($TPP = 0.9$) and high specificity ($TNP = 0.9$). A horizontal dashed line represents the prior probability $\Pr(D_1) = 0.05$, at which $I(T_1) = 0.37$ nits (Equation 6) (we are not concerned with $I(T_2)$ here). The vertical line from $I(T_1) = 0.37$ nits intersects the curve above $\Pr(D_1) = 0.05$ at $\Pr(D_1|T_1) = 0.32$ (as indicated by the horizontal blue line). Note that an expected information content of $I(T_1) \approx 1$ nit is required to raise the posterior probability of need for fungicide application above $\Pr(D_1|T_1) = 0.5$ with the specified $\Pr(D_1)$.

and $h_{ij} > 0$ if the true status $D_j$ occurs more frequently when prediction $T_i$ is made than under independence, and $h_{ij} < 0$ in the opposite case. Thus $h_{ij}$ represents the information content of the prediction $T_i$ in relation to the true status $D_j$ (Equation 2) and in the special case of equal subscripts (i.e., $i=j$), $h_{ij}$ is the information gain (section 2.1).

We now calculate the expected mutual information, denoted here $I_M$. Working in natural logarithms, the expected mutual information in nits is given by:

$$I_M = \sum_{i=1}^{n} \sum_{j=1}^{m} \Pr(T_i \cap D_j) \ln\left[\frac{\Pr(D_j|T_i)}{\Pr(D_j)}\right]$$

(14)

Thus, like relative entropy (section 2.2), expected mutual information is a form of expected information content. Specifically, relative entropy provides us with a measure of the expected information content for a particular prediction, given the prior probability and the resulting posterior probabilities, while expected mutual information refers to expected information content taken over all $n$ possible predictions: expected mutual information is expected relative entropy. Expected mutual information provides a measure of the average amount of information obtained from a predictor, given the prior probability (see Benish, 2003). For the Sclerotinia stem rot predictor with threshold risk score = 50, with prior probability $\Pr(D_1)$=0.16, $I_M$ = 0.048 nits (see also Fig. 4). Despite this low value for the amount of information obtained from the predictor averaged over both categories for predicted status and both categories for true status, it is nevertheless worth noting that the information gain $\ln[\Pr(D_1|T_1)/ \Pr(D_1)]$ is large and that this contributes to a large value for $I(T_1)$, the expected information content of a $T_1$ prediction (Table 3). This suggests that the predictor has particular application if our priority is to reduce uncertainty relating to the need for treatment.

Specifically for the case of $m=2$, $n=2$, we can write:

$$I_M = \Pr(T_1)I(T_1) + \Pr(T_2)I(T_2)$$

from which, using Equations 7 and 9, we obtain:

$$\begin{aligned} I_M &= \Pr(D_1)\Pr(T_1|D_1)\ln[\Pr(T_1|D_1)] + \Pr(D_2)\Pr(T_1|D_2)\ln[\Pr(T_1|D_2)] \\ &+ \Pr(D_1)\Pr(T_2|D_1)\ln[\Pr(T_2|D_1)] + \Pr(D_2)\Pr(T_2|D_2)\ln[\Pr(T_2|D_2)] \\ &- \Pr(T_1)\ln[\Pr(T_1)] - [1 - \Pr(T_1)]\ln[1 - \Pr(T_1)] \end{aligned}$$

(15)

which is equation 7 from Benish (2002) (Benish works in logarithms base 2). Thus for $m=2$, $I_M$ is a function of the prior probability $\Pr(D_1)$ (since $\Pr(D_2) = 1-\Pr(D_1)$) and the conditional probabilities $\Pr(T_1|D_1)$, $\Pr(T_1|D_2)$, $\Pr(T_2|D_1)$ and $\Pr(T_2|D_2)$, which we recognize as, respectively, the true positive proportion (*TPP*, sensitivity), the false positive proportion (*FPP*, 1−specificity), the false negative proportion (*FNP*, 1−sensitivity) and the true negative proportion (*TNP*, specificity) of the predictor (section 1.4). $\Pr(T_1)$ is obtained as in Table 4.

For a perfect predictor, we have $\Pr(D_1|T_1) = \Pr(D_2|T_2) = 1$, $\Pr(D_2|T_1) = \Pr(D_1|T_2) = 0$, and we take $0\ln(0)=0$. Then Equation 15 reduces to:

$$I_M = \Pr(D_1)\ln\left[\frac{1}{\Pr(D_1)}\right] + \Pr(D_2)\ln\left[\frac{1}{\Pr(D_2)}\right]$$

(16)

with $\Pr(D_1) + \Pr(D_2) = 1$. This is Equation 3 for the case of $m=2$.

Equation 15 (and Equation 16) can be used to plot graphs showing relationships between the mutual information and the prior probability (in the present notation) $\Pr(D_1)$ (Benish, 2002). Mutual information is zero when $\Pr(D_1) = 0$ or $\Pr(D_1) = 1$, since we are already certain of the outcome. The intermediate value of $\Pr(D_1)$ at which the $I_M$ value is maximized can be found by differentiating Equation 15, equating the resulting expression to zero, and solving for $\Pr(D_1)$. From Equation 15, the maximum value of $I_M$ occurs at the value of $\Pr(D_1)$ given by:

$$\Pr(D_1) = \frac{TNP \exp(K) - FPP}{J(\exp(K) + 1)} \tag{17}$$

where $\quad K = \dfrac{TPP \ln(TPP) - FPP \ln(FPP) + FNP \ln(FNP) - TNP \ln(TNP)}{J} \quad$ and $J$ is Youden's index, $J = TPP + TNP - 1$. For a perfect predictor, from Equation 16, the maximum value of $I_M$ occurs at the value of $\Pr(D_1) = 0.5$, when $I_M = 0.69$ nits. Fig. 4 shows information graphs for the predictor for Sclerotinia stem rot of oil seed rape described in Example 2. Mutual information is highest when the threshold risk score = 35 and lowest when the threshold risk score = 50.

### 2.5. Expected information and ROC curves

Recall that in Example 2 (section 1.3), different choices of threshold score for the indicator variable resulted in different values for sensitivity and specificity. Suppose now that instead of restricting the choice of threshold to one, two or three values of the indicator score, we allow the threshold indicator score to vary over the whole range of possible indicator scores. Note that it is normally the case that the indicator score is calibrated so as to be positively correlated with the perceived risk. A graphical plot of $TPP$ (sensitivity) against $FPP$ (1−specificity), known as a *receiver operating characteristic* (ROC) curve, provides a useful summary of the characteristics of the indicator in question. ROC curves are widely used in clinical chemistry for the evaluation of diagnostic tests (e.g., Metz, 1978; Zweig and Campbell, 1993; Swets et al., 2000). An overview of ROC curves in an ecological context is provided by Murtaugh (1996). Yuen et al. (1996) and Twengström et al. (1998) have pioneered the phytopathological application of ROC curves.

ROC curve methodology is applicable when the true status of a crop, $D_j$ ($j = 1...m$), is described in one of two categories ($m$=2) and the predicted status, $T_i$ ($i = 1...n$), is also described in one of two categories ($n$=2). As before, $D_1$ denotes that the true status is a disease outbreak, or the need for treatment and $D_2$ denotes that the true status is no outbreak, or no need for treatment. $T_1$ denotes a prediction of a disease outbreak, or the need for treatment and $T_2$ denotes a prediction of no outbreak, or no need for treatment. Here we use an ROC curve of the form:

$$TPP = \left[1 + \exp(-\Delta) \cdot \left(FPP^{-\mu} - 1\right)\right]^{-1/\mu} \tag{18}$$

(Lloyd, 2000) as a basis for some illustrative calculations. Values for parameters $\Delta$ and $\mu$ will be assumed rather than estimated from data.

Fig. 5 shows an ROC plot with a curve based on Equation 18, with $\Delta$=2.4, $\mu$=0.4. Suppose a threshold indicator score is chosen such that all cases and all controls are declared positive (i.e., all the indicator scores are above the threshold). This classification is correct for all the cases, so $TPP = 1$. However, it is wrong for all the controls, so $FPP = 1$. Thus, the corresponding point on the ROC curve is (1,1), in the extreme top right-hand corner of the ROC plot. Now consider a threshold indicator score chosen such that all cases and all controls are declared negative (i.e., all the indicator scores are at or below the threshold). This classification is wrong for all the cases, so $TPP = 0$. However, it is correct for all the controls, so $FPP = 0$. Thus, the corresponding point on the ROC plot is at the origin, (0,0).

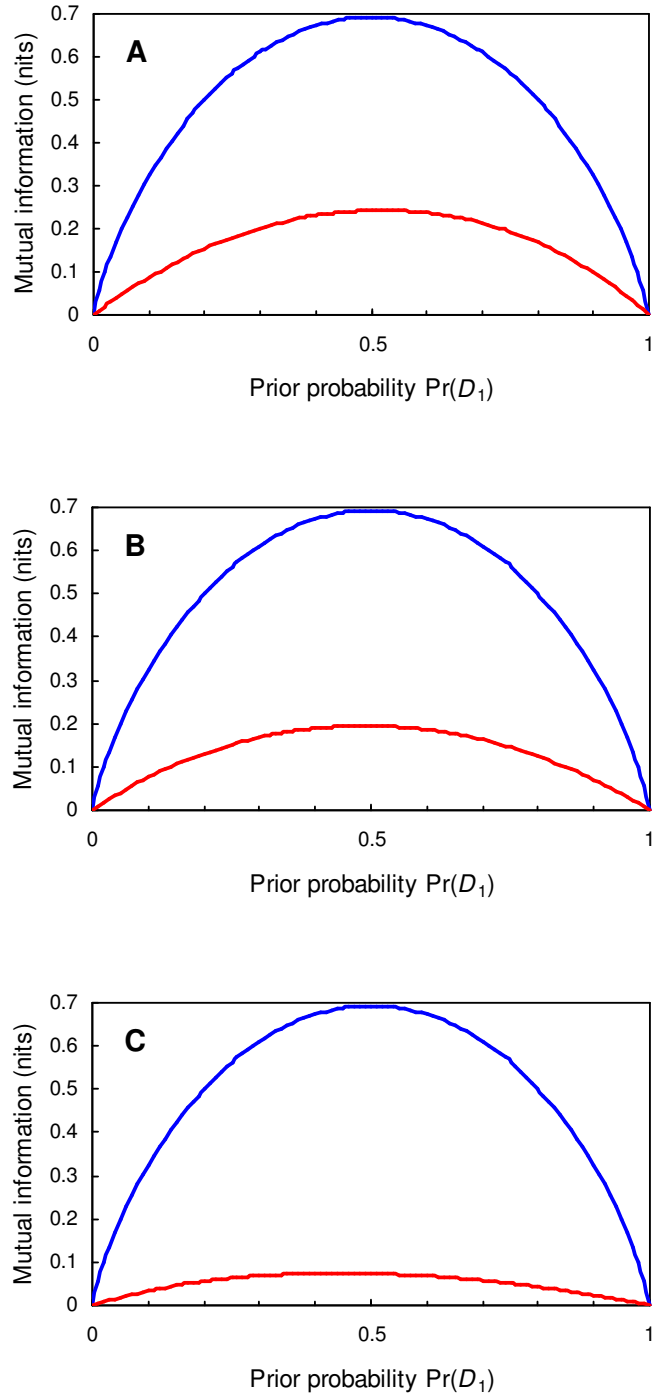

**FIG. 4.** Mutual information graphs ($I_M$ shown in red) for the Sclerotinia stem rot of oil seed rape predictor described in Example 2 (see section 1.3). **A.** Threshold risk score = 35. The maximum value of $I_M = 0.243$ nits occurs at $Pr(D_1) = 0.519$. At $Pr(D_1) = 0.16$, $I_M = 0.13$ nits. **B.** Threshold risk score = 40. The maximum value of $I_M = 0.195$ nits occurs at $Pr(D_1) = 0.49$. At $Pr(D_1) = 0.16$, $I_M = 0.11$ nits. **C.** Threshold risk score = 50. The maximum value of $I_M = 0.075$ nits occurs at $Pr(D_1) = 0.443$. At $Pr(D_1) = 0.16$, $I_M = 0.048$ nits. On each graph is also shown a blue line that represents the mutual information graph for a perfect predictor. For this predictor, the maximum value of $I_M = 0.693$ nits occurs at $Pr(D_1) = 0.5$. At $Pr(D_1) = 0.16$, $I_M = 0.44$ nits.

If it were possible to place a threshold on the indicator score scale such that all the cases are correctly classified ($TPP = 1$) and all the controls are correctly classified ($TNP = 1$, $FPP = 0$), we would have a test that provides perfect discrimination. On the ROC plot this threshold is the point (0,1) in the extreme top left-hand corner of the plot where, simultaneously, sensitivity and specificity are both equal to one.

If the frequency distributions of indicator scores for cases and controls are identical, $TPP = FPP$ no matter what threshold indicator score is chosen. As a result, the corresponding ROC curve follows a straight line (the "no discrimination" line) along the diagonal between the points (1,1) (low threshold, $TPP = FPP = 1$) and (0,0) (high threshold, $TPP = FPP = 0$). In passing, note that an ROC curve that falls below this line suggests that the indicator in question is giving consistently wrong results. Such an indicator has some discriminatory capability insofar as useful predictions could be made by inverting the results obtained by its application.

When there is partial overlap between the distributions of indicator scores for cases and controls, the ROC curve for a useful indicator will exhibit a curvature away from the no discrimination line towards the top left-hand corner of the plot (Fig. 5; see also, for example, Twengström et al. (1998) for the Sclerotinia stem rot data discussed in Example 2). An indicator that has an ROC curve with a curvature towards to the top left-hand corner of the plot has desirable sensitivity and specificity characteristics, in that relatively high values of both can be achieved simultaneously with an appropriate choice of threshold indicator score. For a given choice of threshold, Youden's index $J$ = sensitivity + specificity − 1 (= $TPP − FPP$) (Youden, 1950), characterizes the overall non-error rate of the corresponding test. $J$ is equal to one for a test that enables perfect discrimination and is equal to zero for a test that provides no discrimination.

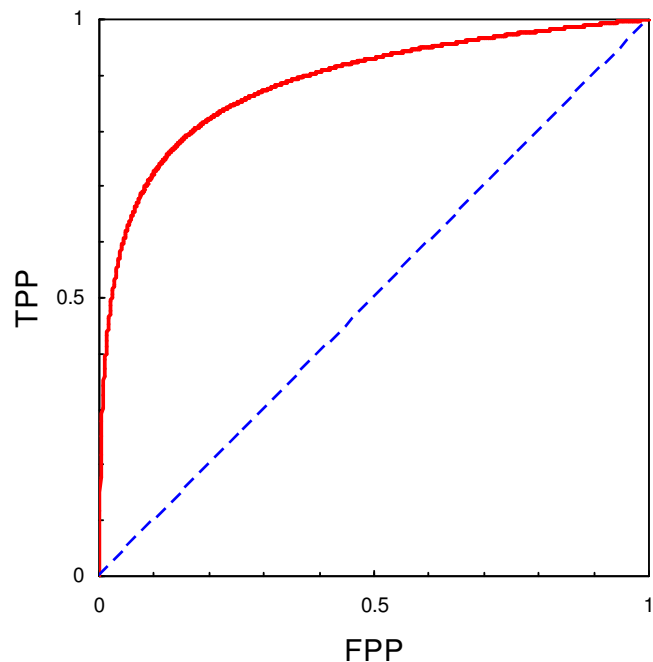

**FIG. 5.** A receiver operating characteristic curve (in red) based on Equation 18, with Δ=2.4, μ=0.4. The dashed blue line along the diagonal between the points (1,1) (low threshold, $TPP = FPP = 1$) and (0,0) (high threshold, $TPP = FPP = 0$) is the "no discrimination" line.

Metz et al. (1973) characterized ROC curves in terms of mutual information. When the true status of a crop, $D_j$ ($j = 1…m$), is described in one of two categories ($m$=2), mutual information ($I_M$) can be expressed in terms of the prior probability $\Pr(D_1)$ and the conditional probabilities $\Pr(T_1|D_2)$, $\Pr(T_1|D_2)$, $\Pr(T_2|D_1)$ and $\Pr(T_2|D_2)$ (Equation 15). $\Pr(T_1|D_1)$ is the true positive proportion (*TPP*, sensitivity) and $\Pr(T_2|D_1)$ is the false negative proportion (*FNP*, 1−sensitivity). $\Pr(T_1|D_2)$ is the false positive proportion (*FPP*, 1−specificity) and $\Pr(T_2|D_2)$ is the true negative proportion (*TNP*=1−*FPP*, specificity). Thus, for a given prior probability, expected mutual information can be expressed in terms of *TPP* and *FPP*. This is shown explicitly in equation 2 of Metz et al. (1973), which can be obtained directly by making the appropriate substitutions in Equation 15 and then rearranging (note that Metz et al. work in logarithms base 2). Now, contours showing the loci of points corresponding to designated values of expected mutual information can be constructed on the graph with *TPP* on the ordinate and *FPP* on the abscissa (for a given prior probability) (Fig. 6). Metz et al. (1973) refer to these contours as *iso-information curves* and give some examples.

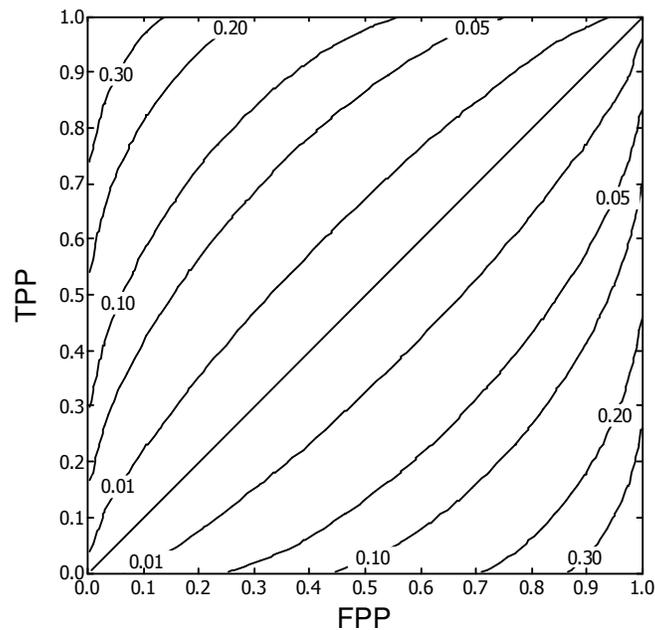

**FIG. 6.** Iso-information contours for expected mutual information ($I_M$) in nits plotted on the graph with *TPP* on the ordinate and *FPP* on the abscissa (i.e., the axes of an ROC curve). Calculations are based on Equation 15, with a prior probability $\Pr(D_1) = 0.2$. Note (1) that the no discrimination line on an ROC plot (see Fig. 5) is identical to the contour for $I_M = 0$ and (2) that contours below $I_M = 0$ conceptualize, in information terms, the idea that an ROC curve that falls below the no discrimination line still has some useful discriminatory capability.

The expected mutual information corresponding to any particular set of (*FPP*,*TPP*) coordinates on an ROC curve may be obtained either by calculation, from Equation 15, or graphically, by comparing the ROC curve with a set of iso-information curves based on an appropriate value of the prior probability. Metz et al. (1973) suggest that this provides a basis for comparing, in terms of expected mutual information, the implications of adopting different operational threshold scores along an ROC curve. We can think of this as an alternative to the analysis, based on Equation 15, presented in section 2.4 (see Fig. 4). Metz et al. (1973) also suggest that the maximum expected mutual information available from an ROC curve provides an objective, quantitative basis for evaluation of the usefulness of an indicator and for comparison of different indicators used in the same diagnostic system.

Somoza and Mossman (1992) took equation 2 of Metz et al. (1973) (i.e., Equation 15 above) and tried to characterize the maximum expected mutual information available from an ROC curve. Working with a class of continuous ROC curves based on the assumption of Normal probability distributions for the indicator scores of both cases and controls, no analytical result was found for the value of $FPP$ at which expected mutual information was at a maximum. Numerical analysis was presented, as also discussed in Somoza et al. (1989) and Somoza and Mossman (1990).

Writing $I$ to denote expected information content generically, Equation 15, with a given (constant) prior probability, can be written in the general format:

$$I = f(TPP, FPP) \tag{19}$$

so that:

$$\frac{\mathrm{d}TPP}{\mathrm{d}FPP} = -\frac{\partial f / \partial FPP}{\partial f / \partial TPP}$$

and for Equation 15 in particular:

$$\frac{\mathrm{d}TPP}{\mathrm{d}FPP} = \left( \frac{1 - \Pr(D_1)}{\Pr(D_1)} \right) \cdot \left[ \frac{\ln\left( \dfrac{FPP}{\Pr(T_1)} \right) - \ln\left( \dfrac{1 - FPP}{\Pr(T_2)} \right)}{\ln\left( \dfrac{1 - TPP}{\Pr(T_2)} \right) - \ln\left( \dfrac{TPP}{\Pr(T_1)} \right)} \right] \tag{20}$$

in which $\Pr(T_1)$ and $\Pr(T_2)$ are calculated as in Table 4. This is equivalent to equation 6 of Somoza and Mossman (1992) (Somoza and Mossman work in logarithms base 2). This analysis tells us that $I_M$ is maximized if the indicator score adopted as the operational threshold is chosen such that the slope of the ROC curve at the point representing the operational threshold is $\dfrac{\mathrm{d}TPP}{\mathrm{d}FPP}$ as given in Equation 20. Unfortunately, this does not provide us with an analytical basis for finding the $(FPP, TPP)$ coordinates of the point representing the operational threshold.

Working numerically, we can set a range of values of $FPP$ in the interval 0,1 and calculate the corresponding values of $TPP$ from Equation 18 (with given values of constants $\Delta$ and $\mu$). Then, for each pair of $(FPP, TPP)$ coordinates, and a given value of the prior probability $\Pr(D_1)$, a value of $I_M$ can be calculated from Equation 15, and the maximum value of $I_M$ found by inspection (Fig. 7). The result can be checked, because we can calculate the slope of the ROC curve at the point where $I_M$ is maximized from the derivative of Equation 18:

$$\frac{\mathrm{d}TPP}{\mathrm{d}FPP} = \left[ 1 + \exp(-\Delta) \cdot FPP^{-\mu} - \exp(-\Delta) \right]^{-(\mu+1)/\mu} \cdot \exp(-\Delta) \cdot FPP^{-(\mu+1)} \tag{21}$$

and it should be the same as the value of the derivative calculated from Equation 20 at the same point.

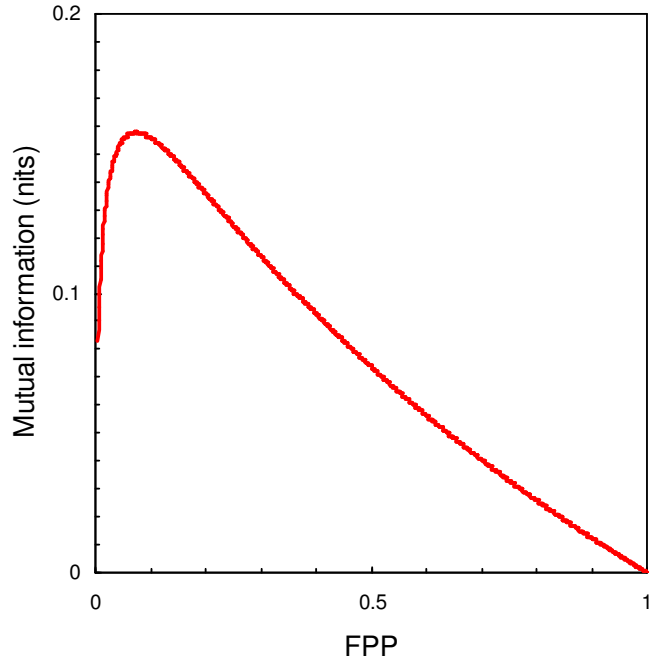

**FIG. 7.** Expected mutual information ($I_M$) in nits calculated from Equation 15, with prior probability $\Pr(D_1) = 0.2$ and values of *FPP* and the corresponding *TPP* taken from the ROC curve in Fig. 5. By inspection, the maximum value of $I_M$ is 0.158 nits, and occurs at the point *FPP* = 0.073, *TPP* = 0.679 on the ROC curve. At this point, the derivative d*TPP*/d*FPP* = 2.06, from both Equations 20 and 21.

For completeness, we now take a similar numerical and graphical approach to the analysis of expected information content (i.e., relative entropy) as described by Equation 7 for $I(T_1)$ (relative entropy for prediction $T_1$) and Equation 9 for $I(T_2)$ (relative entropy for prediction $T_2$). Working numerically with Equation 7, we can set a range of values of *FPP* in the interval 0,1 and calculate the corresponding values of *TPP* from Equation 18 (with given values of constants $\Delta$ and $\mu$). Then, for each pair of (*FPP*,*TPP*) coordinates and a given value of the prior probability $\Pr(D_1)$, a value of $I(T_1)$ can be calculated from Equation 7 (Fig. 8). Using the same procedure, values of $I(T_2)$ can be calculated from Equation 9 corresponding to pairs of (*FPP*,*TPP*) coordinates from an ROC curve described by Equation 18 and a given prior probability (Fig. 8).

We see from Fig. 8 that neither the graphical plot of $I(T_1)$ against *FPP* nor $I(T_2)$ against *FPP* has intermediate maximum point for expected information content. Recall that for a predictor with an operational threshold at the extreme top right-hand corner of the ROC curve (where *TPP*=1, *FPP*=1), every prediction will have the outcome (in the current notation) $T_1$ (since $\Pr(T_1|D_1) = 1$ and $\Pr(T_1|D_2) = 1$). In information terms, we see from Fig. 8 that the expected information content for prediction $T_1$, $I(T_1)$, is zero when *FPP*=1, which is reasonable. $I(T_1)$ increases monotonically as *FPP* decreases towards zero. For a predictor with an operational threshold at the extreme bottom left-hand corner of the ROC curve (where *TPP*=0, *FPP*=0), every prediction will have the outcome (in the current notation) $T_2$ (since $\Pr(T_2|D_1) = 1$ and $\Pr(T_2|D_2) = 1$). In information terms, we see from Fig. 8 that the expected information content for prediction $T_2$, $I(T_2)$, is zero when *FPP*=0, which again is reasonable. $I(T_2)$ increases monotonically as *FPP* increases towards one. From Fig. 8, we see that it is not possible to change *FPP* in such a way as to simultaneously increase both $I(T_1)$ and $I(T_2)$. This is the information equivalent of the trade-off between sensitivity and specificity involved in selecting an appropriate operational threshold on an ROC curve.

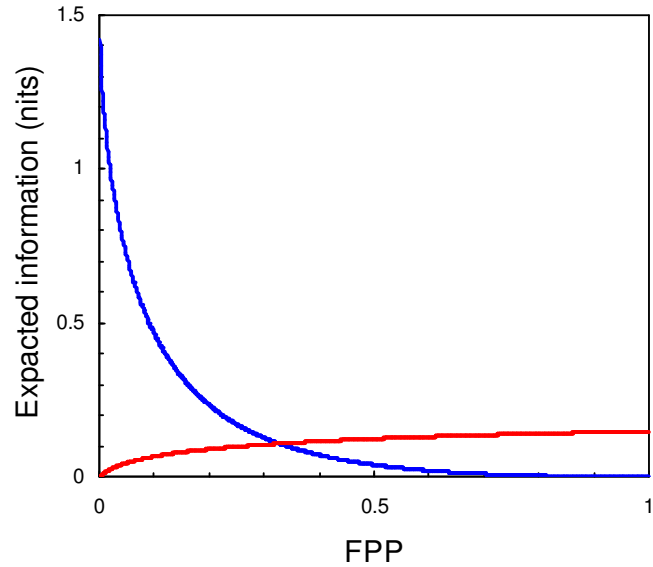

**FIG. 8.** Expected information content. Calculations are based on an ROC curve described by Equation 18, with $\Delta$=2.4, $\mu$=0.4 (Fig. 5), and a prior probability $\Pr(D_1) = 0.2$. $I(T_1)$ in nits (blue line) is calculated from Equation 7, with values of $FPP$ and the corresponding $TPP$ taken from the ROC curve. $I(T_2)$ in nits (red line) is calculated from Equation 9, with values of $FPP$ and the corresponding $TPP$ taken from the ROC curve.

Finally, we note that Equation 7 for $I(T_1)$ and Equation 9 for $I(T_2)$ lead directly to Biggerstaff's (2000) likelihood ratios graph (mentioned in section 1.4). With a given prior probability $\Pr(D_1)$ we obtain, from Equation 7 via Equation 19:

$$\frac{\mathrm{d}TPP}{\mathrm{d}FPP} = \frac{TPP}{FPP}$$

the solution of which is $TPP = aFPP$, in which we recognize $a$ as the likelihood ratio $TPP/FPP$ (section 1.4). From Equation 9 with the same given prior probability $\Pr(D_1)$ we obtain, also via Equation 19:

$$\frac{\mathrm{d}TPP}{\mathrm{d}FPP} = \frac{1-TPP}{1-FPP}$$

the solution of which is $TPP = 1 - b + bFPP$, in which we recognize $b$ as the likelihood ratio $FNP/TNP$ (section 1.4). The two straight line relationships obtained between $TPP$ and $FPP$ are the lines used to construct Biggerstaff's likelihood ratios graph.

## 2.6. Expected information for predictors with multiple outcome categories

Thus far, we have mainly been concerned with the concept of expected information applied to predictors for which the actual status of individuals is described in one of two categories ($m$=2) and the predicted status of individuals is also described in one of two categories (*n*=2). Although predictors of this type are important, there are nevertheless many examples of predictors with multiple outcome categories. In clinical epidemiology, for example, a five-category rating scale for the predicted status of individuals is often adopted (*n*=5), while the actual status of individuals is still described in one of two categories (*m*=2) (see, for example, Swets, 1988). Continuing with the previously introduced notation, the verbal descriptions of the five categories can be written:

$T_1$     'very likely $D_1$',

$T_2$     'probably $D_1$',

$T_3$     'possibly $D_1$',

$T_4$     'probably not $D_1$', and

$T_5$     'very likely not $D_1$'.

Developing such a predictor requires that each of these verbal descriptions is applied to an appropriate range of indicator scores, a more complicated task than adopting a single operational threshold that results in two categories for the predicted status of individuals.

When it comes to characterizing such a predictor, simple concepts of sensitivity and specificity, and their graphical depiction in an ROC curve, are no longer applicable. However, we can generalize Equation 4 as a formula for the expected information content of the *i*th prediction (denoted $I(T_i)$):

$$I(T_i) = \sum_{j=1}^{m} \Pr(D_j|T_i) \ln \left[ \frac{\Pr(D_j|T_i)}{\Pr(D_j)} \right] \qquad (22)$$

and already have an appropriate formula for the expected mutual information in Equation 14. Neither Equation 14 nor Equation 22 are restricted in application to predictors for which *m*=2, *n*=2.

As an example, data for two hypothetical predictors (denoted *A* and *B*) taken from Lee (1999, Table 2) are presented here in the form of numerical prediction-realization tables (Table 5, Table 6). For both predictors, *m* = 2 and *n* = 5, and a prior probability $\Pr(D_1) = 0.2$ has been assumed. For both Tables 5 and 6, part A of the table is a numerical prediction-realization table, with values of $\Pr(T_i \cap D_j)$ in the body of the table, values of $\Pr(T_i)$ in the right-hand margin and values of $\Pr(D_j)$ in the lower margin. Part B of the table has the posterior probabilities $\Pr(D_j|T_i)$, obtained from part A of the table using Equation 5. In the body of part C of the table are the information contents $\ln[\Pr(D_j|T_i)/ \Pr(D_j)]$, obtained from parts A and B using Equation 2. In the right hand margin of part C are the expected information contents $I(T_i)$, obtained using Equation 22. The easiest way to obtain the expected mutual information from the data as presented in Tables 5 and 6 is to calculate $I_M = \sum_{i=1}^{n} \Pr(T_i) I(T_i)$. For predictor *A* (Table 5), $I_M = 0.15$ nits. For predictor *B* (Table 6), $I_M = 0.13$ nits.

Expected mutual information is an overall index of the performance of a predictor, given the prior probability (Benish, 2003). In the current example, the difference in overall performance between predictor *A* and predictor *B* at a prior probability $\Pr(D_1)=0.2$, as characterized by their respective $I_M$ values, is probably nothing to get excited about. Of more interest in this case are the differences in expected information content. Recall that expected information content quantifies the information value of a particular prediction, again given the prior probability (Benish, 1999). We note, for example, that $I(T_1)$ for predictor *A* (Table 5) is rather larger than $I(T_1)$ for predictor *B* (Table 6). On the other hand, $I(T_5)$ for predictor *B* (Table

6) is larger than $I(T_5)$ for predictor $A$ (Table 5). This suggests that if, in making predictions, our priority is to reduce uncertainty relating to the statement that the outcome is 'very likely $D_1$', we should adopt predictor $A$ for use. Alternatively, if our priority is to reduce uncertainty relating to the statement that the outcome is 'very likely not $D_1$', we should instead adopt predictor $B$.

Finally, we note the possibility of calculating expected information as a means of characterizing predictors developed for use in regulatory policy related to risk management for invasive exotic species. Such predictors are often concerned with the identification of potentially harmful weed species among planned imports of exotic plants. In this context, a three-category rating scale may be adopted for both the predicted status of individuals ($n$=3) and the actual status of individuals ($m$=3) (see, for example, Reichard and Hamilton (1997) and Pheloung (1999)). Verbal descriptions of the three categories for the actual status of individuals, often as determined by a consensus of expert opinion, may be written $D_1$ 'serious threat', $D_2$ 'minor threat', and $D_3$ 'non-threat'. Verbal descriptions of the three categories for the outcome of the predictor may be written $T_1$ 'deny entry', $T_2$ 'evaluate further', and $T_3$ 'allow entry'. Characterizing such predictors usually involves combining the 'serious threat' and 'minor threat' categories for the actual status of individuals, and allowing only a 'deny entry' or 'allow entry' decision as the outcome of the predictor. This reduces the problem to one in which the actual status of individuals is described in one of two categories ($m$=2) and the predicted status of individuals is also described in one of two categories ($n$=2). The methodology of ROC curve analysis can then be applied (see, for example, Hughes and Madden (2003) and Caley and Kuhnert (2006)). Use of Equations 14 and 22 to calculate, respectively, the expected mutual information $I_M$ and the expected information contents $I(T_i)$ obviates the need to combine categories for both the predicted status of individuals and the actual status of individuals.

## 2.7. Relative entropy as a measure of diagnostic information

Relative entropy is a synonym for expected information content. As noted by Cover and Thomas (2006), it is sometimes useful to think of relative entropy as a measure of the distance between two distributions. For example, when calculating the expected information content of a prediction (section 2.2), we could think of this as the distance between the distribution of prior probabilities and the distribution of posterior probabilities, given the prediction.

As such, relative entropy can be regarded as a post-prediction measure of expected diagnostic information. In this context, we are interested in calculating the relative entropy of a particular prediction $T_i$, $i = 1…n$, that is one of a set of $n$ predictions that the predictor in question supplies. This relative entropy is a weighted average of the information contents arising from the prediction $T_i$, the average being taken over all $m$ categories of actual status, $D_j$, $j = 1…m$. The weights are the posterior probabilities of the categories of actual status $D_j$, $j = 1…m$, given the particular prediction $T_i$. Benish's (1999) analysis of relative entropy relates to the question: "How much does the prediction $T_i$ tell us, on average, over all $m$ categories of actual status, $D_j$, $j = 1…m$?"

**Table 5.** Data from Test *A* as described in Lee (1999, Table 2), with an assumed prior probability $\Pr(D_1)=0.2$. A. The prediction-realization table. B. Posterior probabilities. C. Information contents.

A.

| Prediction, $T_i$ | Realization, $D_j$ | | Row sums |
|---|---|---|---|
| | $D_1$ | $D_2$ | $\Pr(T_i)$ |
| $T_1$ (very likely $D_1$) | 0.1172 | 0.0432 | 0.1604 |
| $T_2$ (probably $D_1$) | 0.0368 | 0.1760 | 0.2128 |
| $T_3$ (possibly $D_1$) | 0.0260 | 0.1824 | 0.2084 |
| $T_4$ (probably not $D_1$) | 0.0108 | 0.1936 | 0.2044 |
| $T_5$ (very likely not $D_1$) | 0.0092 | 0.2048 | 0.2140 |
| Column sums $\Pr(D_j)$ | 0.2000 | 0.8000 | Grand sum = 1 |

B.

| Prediction, $T_i$ | Posterior probability | | Row sums |
|---|---|---|---|
| | $\Pr(D_1|T_i)$ | $\Pr(D_2|T_i)$ | |
| $T_1$ (very likely $D_1$) | 0.7307 | 0.2693 | 1 |
| $T_2$ (probably $D_1$) | 0.1729 | 0.8271 | 1 |
| $T_3$ (possibly $D_1$) | 0.1248 | 0.8752 | 1 |
| $T_4$ (probably not $D_1$) | 0.0528 | 0.9472 | 1 |
| $T_5$ (very likely not $D_1$) | 0.0430 | 0.9570 | 1 |

C.

| Prediction, $T_i$ | Information content (nits) | | Expected information (nits) |
|---|---|---|---|
| | $\mathrm{Ln}[\Pr(D_1|T_i)/\Pr(D_1)]$ | $\mathrm{Ln}[\Pr(D_2|T_i)/\Pr(D_2)]$ | $I(T_i)$ |
| $T_1$ (very likely $D_1$) | 1.2956 | -1.0887 | 0.6535 |
| $T_2$ (probably $D_1$) | -0.1454 | 0.0333 | 0.0024 |
| $T_3$ (possibly $D_1$) | -0.4719 | 0.0899 | 0.0198 |
| $T_4$ (probably not $D_1$) | -1.3311 | 0.1689 | 0.0896 |
| $T_5$ (very likely not $D_1$) | -1.5373 | 0.1792 | 0.1054 |

**Table 6.** Data from Test $B$ as described in Lee (1999, Table 2), with an assumed prior probability $\Pr(D_1)=0.2$. A. The prediction-realization table. B. Posterior probabilities. C. Information contents.

A.

| Prediction, $T_i$ | Realization, $D_j$ | | Row sums |
|---|---|---|---|
| | $D_1$ | $D_2$ | $\Pr(T_i)$ |
| $T_1$ (very likely $D_1$) | 0.0520 | 0.0336 | 0.0856 |
| $T_2$ (probably $D_1$) | 0.0492 | 0.0480 | 0.0972 |
| $T_3$ (possibly $D_1$) | 0.0444 | 0.1056 | 0.1500 |
| $T_4$ (probably not $D_1$) | 0.0424 | 0.1360 | 0.1784 |
| $T_5$ (very likely not $D_1$) | 0.0120 | 0.4768 | 0.4888 |
| Column sums $\Pr(D_j)$ | 0.2000 | 0.8000 | Grand sum = 1 |

B.

| Prediction, $T_i$ | Posterior probability | | Row sums |
|---|---|---|---|
| | $\Pr(D_1|T_i)$ | $\Pr(D_2|T_i)$ | |
| $T_1$ (very likely $D_1$) | 0.6075 | 0.3925 | 1 |
| $T_2$ (probably $D_1$) | 0.5062 | 0.4938 | 1 |
| $T_3$ (possibly $D_1$) | 0.2960 | 0.7040 | 1 |
| $T_4$ (probably not $D_1$) | 0.2377 | 0.7623 | 1 |
| $T_5$ (very likely not $D_1$) | 0.0245 | 0.9755 | 1 |

C.

| Prediction, $T_i$ | Information content (nits) | | Expected information (nits) |
|---|---|---|---|
| | $\mathrm{Ln}[\Pr(D_1|T_i)/\Pr(D_1)]$ | $\mathrm{Ln}[\Pr(D_2|T_i)/\Pr(D_2)]$ | $I(T_i)$ |
| $T_1$ (very likely $D_1$) | 1.1110 | -0.7120 | 0.3954 |
| $T_2$ (probably $D_1$) | 0.9286 | -0.4824 | 0.2318 |
| $T_3$ (possibly $D_1$) | 0.3920 | -0.1278 | 0.0260 |
| $T_4$ (probably not $D_1$) | 0.1726 | -0.0482 | 0.0042 |
| $T_5$ (very likely not $D_1$) | -2.0976 | 0.1983 | 0.1419 |

### 3. Information for discrimination between cases and controls

The idea that the relative entropy can be thought of as a measure of the distance between two distributions (Cover and Thomas, 2006) provides a useful basis for summarizing its application in characterizing predictors. Provided we continue to bear in mind (as will soon become obvious) that relative entropy is not a true distance, the idea of relative entropy as a distance between distributions is indeed a useful one. As discussed by Benish (1999), relative entropy is a measure of the distance between the distribution of posterior probabilities and the distribution of prior probabilities. As discussed by Lee (1999), relative entropy is a measure of the distance between the distributions of risk scores for cases and controls.

Lee (1999) pointed out that in characterizing and comparing predictors in which there are two categories of true status (denoted here $D_1$ (cases) and $D_2$ (controls)) and $n$ categories of prediction, $T_i$ ($i = 1 \ldots n$), the relative entropies:

$$I(D_1, D_2) = \sum_{i=1}^{n} \Pr(T_i | D_1) \log \left[ \frac{\Pr(T_i | D_1)}{\Pr(T_i | D_2)} \right] \tag{23}$$

and

$$I(D_2, D_1) = \sum_{i=1}^{n} \Pr(T_i | D_2) \log \left[ \frac{\Pr(T_i | D_2)}{\Pr(T_i | D_1)} \right] \tag{24}$$

are useful descriptions of the distance between the discrete distributions of risk score categories for cases and controls, respectively $\Pr(T_i | D_1)$ and $\Pr(T_i | D_2)$ ($i = 1 \ldots n$). We note that $I(D_1, D_2)$ and $I(D_2, D_1)$ are non-negative, are zero if and only if $\Pr(T_i | D_1) = \Pr(T_i | D_2)$ for $i = 1 \ldots n$, and that $I(D_1, D_2)$ is not, in general, equal to $I(D_2, D_1)$ (which is one reason why relative entropy is not a true distance). We will denote the distribution that appears in the numerator of the quotient on the right hand side in Equations 23 and 24 as the *comparison distribution* and the distribution that appears in the denominator of that quotient as the *reference distribution*[3]. Thus, in Equation 23, the distribution of predictor outcomes for cases is the comparison distribution and the distribution of predictor outcomes for controls is the reference distribution. In Equation 24, the distribution of predictor outcomes for controls is the comparison distribution and the distribution of predictor outcomes for cases is the reference distribution. As previously, the choice of base of logarithm is immaterial apart from defining the units of information (section 2.1), and we continue to work in natural logarithms (as did Lee, 1999).

#### 3.1. Lee's analysis of a 2×2 decision table

Table 7 shows, in notational format, the 2×2 decision table that summarizes the characteristics of a predictor such as the one for Sclerotinia stem rot of oil seed rape (Example 2). With numerical values for the quantities in the body of Table 7 it is straightforward to calculate $I(D_1, D_2)$ and $I(D_2, D_1)$ from Equations 23 and 24, respectively. The examples here are taken from Lee (1999). First, a predictor with $\Pr(T_1 | D_1) = 0.8$, $\Pr(T_2 | D_2) = 0.9$ has $I(D_1, D_2) = 1.36$ nits and $I(D_2, D_1) = 1.15$ nits. Second, a predictor with $\Pr(T_1 | D_1) = 0.9$, $\Pr(T_2 | D_2) = 0.8$ has $I(D_1, D_2) = 1.15$ nits and $I(D_2, D_1) = 1.36$ nits.

---

[3] Benish's application of relative entropy did not require explicit specification of a comparison distribution and a reference distribution because there is, for practical purposes, only one relative entropy calculation that is of interest in that case, where the posterior probabilities are the comparison distribution and the prior probabilities the reference distribution.

**Table 7.** The decision table for a predictor with two categories of true status $D_j$, $j=1..m$, $m=2$, and two categories of predicted status, $T_i$, $i=1..n$, $n=2$. In the body of the table, $\Pr(T_1|D_1)$ is the true positive proportion (*TPP*, sensitivity), $\Pr(T_2|D_1)$ is the false negative proportion (*FNP*, 1−sensitivity), $\Pr(T_1|D_2)$ is the false positive proportion (*FPP*, 1−specificity) and $\Pr(T_2|D_2)$ is the true negative proportion ($TNP$=1−$FPP$, specificity).

| Prediction, $T_i$ | True status, $D_j$ | |
|---|---|---|
| | $D_1$ | $D_2$ |
| $T_1$ | $\Pr(T_1|D_1)$ | $\Pr(T_1|D_2)$ |
| $T_2$ | $\Pr(T_2|D_1)$ | $\Pr(T_2|D_2)$ |
| Column sums | 1 | 1 |

In order to interpret these results, we turn to Kullback (1968). From Bayes' theorem, we can write:

$$\ln\left[\frac{\Pr(T_1|D_1)}{\Pr(T_1|D_2)}\right] = \ln\left[\frac{\Pr(D_1|T_1)}{\Pr(D_2|T_1)}\right] - \ln\left[\frac{\Pr(D_1)}{\Pr(D_2)}\right] \tag{25}$$

in which $\Pr(D_2) = 1 - \Pr(D_1)$, $\Pr(D_2|T_1) = 1 - \Pr(D_1|T_1)$, and $\Pr(T_1|D_1)/\Pr(T_1|D_2)$ is the likelihood ratio denoted $LR_1$ (section 1.4). $\text{Ln}(LR_1)$ is a measure of the difference in the logarithm of the odds in favour of $D_1$ after the prediction $T_1$ and before the prediction. For any useful predictor, $\ln(LR_1)>0$ (i.e., $LR_1$>1), so that the prediction $T_1$ increases the odds in favour of $D_1$. Similarly:

$$\ln\left[\frac{\Pr(T_2|D_1)}{\Pr(T_2|D_2)}\right] = \ln\left[\frac{\Pr(D_1|T_2)}{\Pr(D_2|T_2)}\right] - \ln\left[\frac{\Pr(D_1)}{\Pr(D_2)}\right] \tag{26}$$

with $\Pr(D_2|T_2) = 1 - \Pr(D_1|T_2)$, and $\Pr(T_2|D_1)/\Pr(T_2|D_2)$ is the likelihood ratio denoted $LR_2$ (section 1.4). $\text{Ln}(LR_2)$ is a measure of the difference in the logarithm of the odds in favour of $D_1$ after the prediction $T_2$ and before the prediction. For a useful predictor, $\ln(LR_2)<0$ (i.e., $LR_2$<1), so that the prediction $T_2$ decreases the odds in favour of $D_1$.

Now $\ln(LR_1)$ may be thought of as the information resulting from the prediction $T_1$ for discrimination in favour of $D_1$ against $D_2$, and $\ln(LR_2)$ as the information resulting from the prediction $T_2$ for discrimination in favour of $D_1$ against $D_2$. The mean information for discrimination in favour of $D_1$ against $D_2$ is then the weighted average of $\ln(LR_1)$ and $\ln(LR_2)$, the weights being the probabilities of predictions $T_1$ and $T_2$, respectively, among individuals that are actually $D_1$. This is $I(D_1,D_2)$ as given by Equation 23 with $n = 2$.

Using a similar line of argument, it can be shown that $\ln(1/LR_1)$ is a measure of the difference in the logarithm of the odds in favour of $D_2$ after the prediction $T_1$ and before the prediction. $\text{Ln}(1/LR_1)$ may be thought of as the information resulting from the prediction $T_1$ for discrimination in favour of $D_2$ against $D_1$. $\text{Ln}(1/LR_2)$ is a measure of the difference in the logarithm of the odds in favour of $D_2$ after the prediction $T_2$ and before the prediction. $\text{Ln}(1/LR_2)$ may be thought of as the information resulting from the prediction $T_2$ for discrimination in favour of $D_2$ against $D_1$. The mean information for discrimination in favour of $D_2$ against $D_1$ is then the weighted average of $\ln(1/LR_1)$ and $\ln(1/LR_2)$, the weights being the probabilities of predictions $T_1$ and $T_2$, respectively, among individuals that are actually $D_2$. This is $I(D_2,D_1)$ as given by Equation 24 with $n = 2$.

For a predictor with $I(D_1,D_2) > I(D_2,D_1)$ (for example, Lee's (1999) predictor with ($\Pr(T_1|D_1)$ = 0.8, $\Pr(T_2|D_2)$ = 0.9) there is, on average, more information for discrimination in favour of $D_1$ (cases) against $D_2$ (controls) than for $D_2$ against $D_1$. For a predictor with $I(D_2,D_1) > I(D_1,D_2)$ (for example, Lee's (1999) predictor with ($\Pr(T_1|D_1)$ = 0.9, $\Pr(T_2|D_2)$ = 0.8) there is, on average, more information for discrimination in favour of $D_2$ (controls) against $D_1$ (cases) than for $D_1$ against $D_2$. Thus, in this particular example, we could adopt the predictor with ($\Pr(T_1|D_1)$ = 0.8, $\Pr(T_2|D_2)$ = 0.9 if our priority is to reduce uncertainty relating to the identification of cases. However, if our priority is to reduce uncertainty relating to the identification of controls, we could adopt the predictor with ($\Pr(T_1|D_1)$ = 0.9, $\Pr(T_2|D_2)$ = 0.8. Note that this analysis characterizes the two predictors on the basis of their respective distributions of predictor outcomes for cases and controls, independent of the prior probability.

Equations 23 and 24 allow construction of iso-information curves showing the loci of points corresponding to designated values of, respectively, $I(D_1,D_2)$ (Fig. 9) and $I(D_2,D_1)$ (Fig. 10) on the graph with $\Pr(T_1|D_1)$ (i.e., $TPP$) on the ordinate and $\Pr(T_1|D_2)$ (i.e., $FPP$) on the abscissa. We can see the contours for both $I(D_1,D_2)$ = 0 (Fig. 9) and $I(D_2,D_1)$ = 0 (Fig. 10) correspond to the no discrimination line on an ROC plot (see Fig. 5). Contours for both $I(D_1,D_2)$ > 0 and $I(D_2,D_1)$ > 0 *below* the no discrimination line on an ROC plot reflect the fact that an indicator with an ROC curve below the no discrimination line has some discriminatory capability, as discussed in section 2.5.

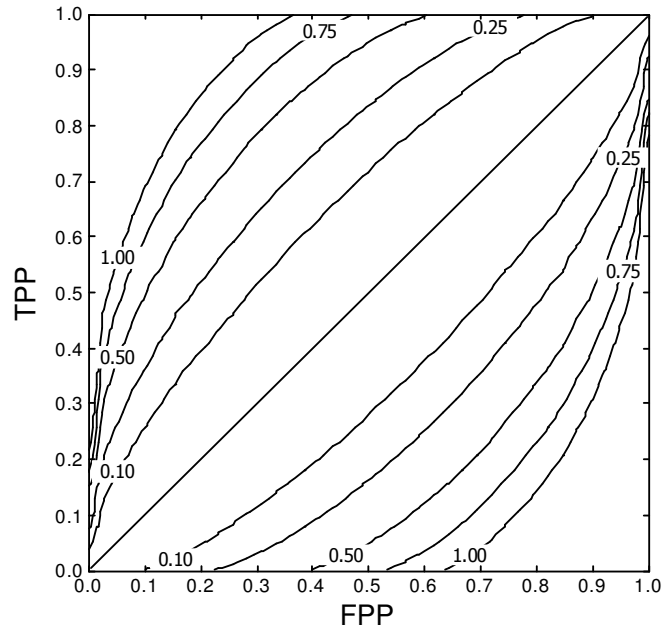

**FIG. 9.** Iso-information contours for expected information $I(D_1,D_2)$ in nits plotted on the graph with $TPP$ on the ordinate and $FPP$ on the abscissa (i.e., the axes of an ROC curve). Calculations are based on Equation 23.

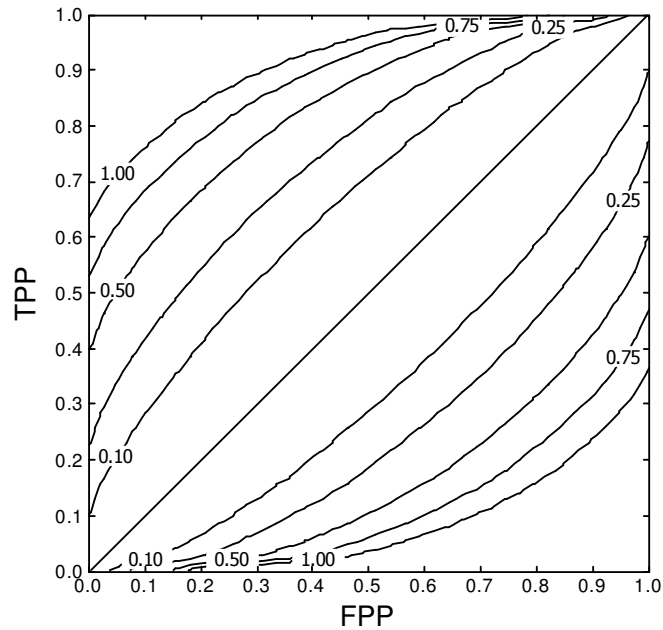

**FIG. 10.** Iso-information contours for expected information $I(D_2, D_1)$ in nits plotted on the graph with *TPP* on the ordinate and *FPP* on the abscissa (i.e., the axes of an ROC curve). Calculations are based on Equation 24.

Finally, in this section, let us write out Equations 23 and 24 explicitly in terms of the ROC curve variables *TPP* and *FPP*:

$$I(D_1, D_2) = TPP \cdot \ln\left[\frac{TPP}{FPP}\right] + (1 - TPP) \cdot \ln\left[\frac{1 - TPP}{1 - FPP}\right] \tag{27}$$

and:

$$I(D_2, D_1) = (1 - FPP) \cdot \ln\left[\frac{1 - FPP}{1 - TPP}\right] + FPP \cdot \ln\left[\frac{FPP}{TPP}\right] \tag{28}$$

Both Equations 27 and 28 have the same generic format as Equation 19, and we can therefore obtain:

$$\frac{\mathrm{d}TPP}{\mathrm{d}FPP} = \frac{\left(\frac{TPP}{FPP}\right) - \left(\frac{1 - TPP}{1 - FPP}\right)}{\ln\left(\frac{TPP}{FPP}\right) - \ln\left(\frac{1 - TPP}{1 - FPP}\right)} \tag{29}$$

from Equation 27 and:

$$\frac{\mathrm{d}TPP}{\mathrm{d}FPP} = \frac{\ln\left(\frac{1 - FPP}{1 - TPP}\right) - \ln\left(\frac{FPP}{TPP}\right)}{\left(\frac{1 - FPP}{1 - TPP}\right) - \left(\frac{FPP}{TPP}\right)} \tag{30}$$

from Equation 28. This analysis tells us that $I(D_1,D_2)$ is maximized if the indicator score adopted as the operational threshold is chosen such that the slope of the ROC curve at the point representing the operational threshold is $\dfrac{\mathrm{d}TPP}{\mathrm{d}FPP}$ as given in Equation 29. $I(D_2,D_1)$ is maximized if the indicator score adopted as the operational threshold is chosen such that the slope of the ROC curve at the point representing the operational threshold is $\dfrac{\mathrm{d}TPP}{\mathrm{d}FPP}$ as given in Equation 30.

Working numerically with an ROC curve described by Equation 18, as before, we can set a range of values of $FPP$ in the interval 0,1 and calculate the corresponding values of $TPP$ from Equation 18 (with given values of constants $\Delta$ and $\mu$). Then, for each pair of $(FPP,TPP)$ coordinates, a value of $I(D_1,D_2)$ can be calculated from Equation 27, and a value of $I(D_2,D_1)$ from Equation 28. The maximum values of $I(D_1,D_2)$ and $I(D_2,D_1)$ can then be found by inspection (Fig. 11). The results can be checked, because we can use Equation 21 to calculate the slopes of the ROC curve at the points where $I(D_1,D_2)$ and $I(D_2,D_1)$ are maximized, and they should be the same as the values of the respective derivatives calculated from Equation 29 and 30 at the same points.

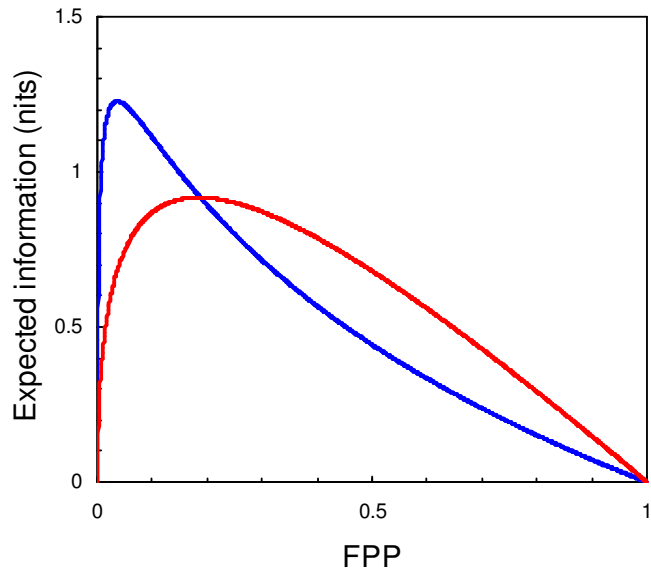

**FIG. 11.** Expected information. Calculations are based on an ROC curve described by Equation 18, with $\Delta=2.4$, $\mu=0.4$ (Fig. 5). $I(D_1,D_2)$ in nits (blue line) is calculated from Equation 27, with values of $FPP$ and the corresponding $TPP$ taken from the ROC curve. By inspection, the maximum value of $I(D_1,D_2)$ is 1.23 nits, and occurs at the point $FPP = 0.037$, $TPP = 0.574$ on the ROC curve. At this point, the derivative $\mathrm{d}TPP/\mathrm{d}FPP = 4.2$, from both Equations 29 and 21. $I(D_2,D_1)$ in nits (red line) is calculated from Equation 28, with values of $FPP$ and the corresponding $TPP$ taken from the ROC curve. By inspection, the maximum value of $I(D_2,D_1)$ is 0.92 nits, and occurs at the point $FPP = 0.184$, $TPP = 0.810$ on the ROC curve. At this point, the derivative $\mathrm{d}TPP/\mathrm{d}FPP = 0.72$, from both Equations 30 and 21.

### 3.2. Interval likelihood ratios

Consider a predictor for which there are two categories of true status ($D_1$ (cases) and $D_2$ (controls)) and $n$ categories of prediction, $T_i$ ($i = 1…n$, $n \geq 2$). Likelihood ratios $LR_i$ are calculated as follows:

$$LR_i = \frac{\Pr(T_i | D_1)}{\Pr(T_i | D_2)} \tag{31}$$

and also, where required:

$$\frac{1}{LR_i} = \frac{\Pr(T_i | D_2)}{\Pr(T_i | D_1)} \tag{32}$$

($i = 1…n$, $n \geq 2$).

When $n = 2$, it is common practice to adopt the terminology of ROC curve methodology and use the true positive proportion (*TPP*, sensitivity) as an estimate of $\Pr(T_1 | D_1)$, the false negative proportion (*FNP*, 1−sensitivity) as an estimate of $\Pr(T_2 | D_1)$, the false positive proportion (*FPP*, 1−specificity) as an estimate of $\Pr(T_1 | D_2)$ and the true negative proportion (*TNP*=1−*FPP*, specificity) as an estimate of $\Pr(T_2 | D_2)$. Then we can write $LR_1 = TPP/FPP =$ sensitivity/(1−specificity) and $LR_2 = FNP/TNP = $ (1−sensitivity)/specificity.

When $n > 2$ (i.e., there are more than two categories of prediction outcome), ROC curve terminology is no longer applicable. The quantities sensitivity (*TPP*) and specificity (*TNP*) cannot be calculated when prediction outcomes cannot be categorized as 'positive' or 'negative'. However, likelihood ratios can still be calculated from Equation 31. These are often called *interval likelihood ratios*. If the categories denoted by $i = 1…n$ are translated into intervals on the original measurement scale of the indicator variable, the $LR_i$ together constitute a discrete *likelihood ratio function*. Although this function is the ratio of two probability mass functions (see Equation 31) it is not, in general, a probability mass function itself.

### 3.3. Lee's analysis of an $n \times 2$ decision table

Table 8 gives details of two hypothetical predictors taken from Lee (1999, Table 2). In each case there are five categories of prediction outcome, and therefore five interval likelihood ratios. The relative entropies $I(D_1,D_2)$ and $I(D_2,D_1)$ can be calculated from Equations 23 and 24, respectively. $I(D_1,D_2)$ is the weighted average of the $\ln(LR_i)$ values, the weights being the distribution of categories of predictor outcomes for cases. $I(D_2,D_1)$ is the weighted average of the $\ln(1/LR_i)$ values, the weights being the distribution of categories of predictor outcomes for controls. Thus, the two relative entropies provide a concise summary of the characteristics of a predictor (of which more in section 3.4).

From Table 8, for predictor $A$, $I(D_1,D_2) = 1.13$ nits (Equation 23) and $I(D_2,D_1) = 0.84$ nits (Equation 24). For predictor $B$, $I(D_1,D_2) = 0.85$ nits (Equation 23) and $I(D_2,D_1) = 1.10$ nits (Equation 24). For predictor $A$, $I(D_1,D_2) > I(D_2,D_1)$ so there is, on average, more information for discrimination in favour of $D_1$ (cases) against $D_2$ (controls) than for $D_2$ against $D_1$. For predictor $B$, $I(D_2,D_1) > I(D_1,D_2)$ so there is, on average, more information for discrimination in favour of $D_2$ (controls) against $D_1$ (cases) than for $D_1$ against $D_2$. This suggests that we should adopt predictor A if our priority is to reduce uncertainty relating to the identification of cases, and predictor B if our priority is to reduce uncertainty relating to the identification of controls.

**Table 8.** The decision tables for two predictors taken from Lee (1999, Table 2). Each predictor has two categories of true status $D_j$, $j=1..m$, $m=2$, and five categories of predicted status, $T_i$, $i=1..n$, $n=5$. In the body of each part of the table, the 'true status' is the conditional probability $\Pr(T_i|D_j)$.

Predictor $A$.

| Prediction, $T_i$ | True status, $D_j$ | | $LR_i$ | $1/LR_i$ |
|---|---|---|---|---|
| | $D_1$ | $D_2$ | (Eq. 31) | (Eq. 32) |
| $T_1$ (very likely $D_1$) | 0.586 | 0.054 | 10.85 | 0.09 |
| $T_2$ (probably $D_1$) | 0.184 | 0.220 | 0.84 | 1.19 |
| $T_3$ (possibly $D_1$) | 0.130 | 0.228 | 0.57 | 1.75 |
| $T_4$ (probably not $D_1$) | 0.054 | 0.242 | 0.22 | 4.48 |
| $T_5$ (very likely not $D_1$) | 0.046 | 0.256 | 0.18 | 5.57 |
| | Sum=1 | Sum=1 | | |

Predictor $B$.

| Prediction, $T_i$ | True status, $D_j$ | | $LR_i$ | $1/LR_i$ |
|---|---|---|---|---|
| | $D_1$ | $D_2$ | (Eq. 31) | (Eq. 32) |
| $T_1$ (very likely $D_1$) | 0.260 | 0.042 | 6.19 | 0.16 |
| $T_2$ (probably $D_1$) | 0.246 | 0.060 | 4.10 | 0.24 |
| $T_3$ (possibly $D_1$) | 0.222 | 0.132 | 1.68 | 0.59 |
| $T_4$ (probably not $D_1$) | 0.212 | 0.170 | 1.25 | 0.80 |
| $T_5$ (very likely not $D_1$) | 0.060 | 0.596 | 0.10 | 9.93 |
| | Sum=1 | Sum=1 | | |

### 3.4. Relative entropy as a weighted average log likelihood ratio

As discussed by Lee (1999), the relative entropy is a measure of the distance between the distribution of categories of prediction outcome for cases (those definitively classified $D_1$) and the distribution of categories of prediction outcome for controls (those definitively classified $D_2$). In this context, neither of these distributions distribution is obviously identifiable as either the comparison distribution or the reference distribution. Thus two relative entropies are calculated, one with the distribution of categories of prediction outcome for cases as the comparison distribution and the distribution of categories of prediction outcome for controls as the reference distribution, the other vice versa. The former can be regarded as a measure of expected information available for discrimination in favour of $D_1$ against $D_2$, while the latter is a measure of expected information for discrimination in favour of $D_2$ against $D_1$. We do not need to know the prior probabilities in order to calculate these two relative entropies: they are characteristics of a predictor independent of $\Pr(D_1)$ and $\Pr(D_2)$. The fact that, in general, these two relative entropies do not have the same value reminds us that thinking of relative entropy as a distance is just a device to help in understanding what otherwise may seem a rather abstract concept.

So, first we consider cases (individuals actually $D_1$) as the comparison distribution: such individuals may have any of $n$ predicted outcomes, $T_i$, $i = 1…n$. For any particular prediction $T_i$, we are interested in the change in the logarithm of the odds of $D_1$ before the prediction and after. This is characterized by the logarithm of the likelihood ratio, denoted here $\ln(LR_i)$. The relative entropy is the weighted average of the logarithm of the likelihood ratio, the average being taken over all $n$ categories of prediction that the predictor in question supplies, $T_i$, $i = 1\ldots n$. The weights are provided by the distribution of prediction outcomes for actually $D_1$ individuals.

We now consider controls (individuals actually $D_2$) as the comparison distribution: such individuals may also have any of $n$ predicted outcomes, $T_i$, $i = 1\ldots n$. For any particular prediction $T_i$, we are interested in the change in the logarithm of the odds of $D_2$ before the prediction and after. This is characterized by the logarithm of the reciprocal of the likelihood ratio, denoted $\ln(1/LR_i)$. In this case, the required relative entropy is the weighted average of the logarithm of the reciprocal of the likelihood ratio, the average being taken over all $n$ categories of prediction that the predictor in question supplies, $T_i$, $i = 1\ldots n$. The weights are provided by the distribution of prediction outcomes for actually $D_2$ individuals.

Lee's (1999) analysis of relative entropy is essentially placed in the context of the questions:

- "How much does the use of a predictor tell us, on average over all $n$ categories of prediction $T_i$, $i = 1\ldots n$, about $D_1$ individuals?" (in which case we have the distribution of categories of prediction outcome for cases as the comparison distribution and the distribution of categories of prediction outcome for controls as the reference distribution) and

- "How much does the use of a predictor tell us, on average over all $n$ categories of prediction $T_i$, $i = 1\ldots n$, about $D_2$ individuals?" (in which case we have the distribution of categories of prediction outcome for controls as the comparison distribution and the distribution of categories of prediction outcome for cases as the reference distribution).

The two relative entropies discussed by Lee (1999) are particularly useful for summarizing the characteristics of predictors with $n>2$ categories of predicted outcome, when sensitivity and specificity as used in ROC curve analysis cannot be calculated.

## 4.    The information properties of indicator variables

The development of a predictor (i.e., a diagnostic test) requires at the outset the characterization of an appropriate indicator variable. Somoza et al. (1989) provide a useful summary of the methodology. Two mutually exclusive groups of subjects are identified, one of crops designated cases, definitively having disease status $D_1$, the other of crops designated controls, definitively having disease status $D_2$. The classification into case and control groups is made independent of the putative indicator variable. The value of this indicator variable is recorded for all subjects in both groups, and the distributions of indicator scores plotted separately for cases and controls. Generally, the two distributions of indicator scores overlap. The indicator variable is usually calibrated in such a way that cases tend to have larger indicator scores than controls.

### 4.1.  Properties of the distributions of indicator scores

Here, we are interested in the information properties of the indicator variable as characterized by the distributions of indicator scores for cases and controls. For the purpose of illustration, we assume that the observed indicator scores are realizations of a variable $X$ that is inherently continuous, and further, that the underlying distribution of indicator scores is Normal (after suitable transformation, if necessary) for both cases and controls (Fig. 12).

For cases we write the probability density function $f_1(x) \sim N(\mu_1, \sigma_1)$, and for controls we write $f_2(x) \sim N(\mu_2, \sigma_2)$, with $\mu$ and $\sigma$ symbolising, respectively, the underlying mean and standard deviation. For continuous distributions, we can calculate the relative entropies $I(f_1, f_2)$ (with cases as the comparison distribution and controls as the reference distribution as in Equation 23, the corresponding discrete version):

$$I(f_1, f_2) = \int_X f_1(x) \log\left[\frac{f_1(x)}{f_2(x)}\right] dx \tag{33}$$

and $I(f_2, f_1)$ (with controls as the comparison distribution and cases as the reference distribution as in Equation 24, the corresponding discrete version):

$$I(f_2, f_1) = \int_X f_2(x) \log\left[\frac{f_2(x)}{f_1(x)}\right] dx \tag{34}$$

The continuous distribution form of relative entropy, referred to as *differential relative entropy*, shares all the properties of the corresponding discrete form and is the limiting form where $f_1(x)$ and $f_2(x)$ are the probability distributions corresponding, respectively, with $\Pr(T_i|D_1)$ and $\Pr(T_i|D_2)$ as discretized into categories of prediction outcomes $T_i$ ($i = 1 \ldots n$; $n \rightarrow \infty$) (Kleeman, 2006).

In numerical calculations, we will continue to work in natural logarithms. Now, specifically for $f_1(x)$ and $f_2(x)$ both Normal, we can write:

$$I(f_1, f_2) = \frac{1}{2}\left[\ln\left(\frac{\sigma_2^2}{\sigma_1^2}\right) - 1 + \frac{\sigma_1^2}{\sigma_2^2} + \frac{(\mu_1 - \mu_2)^2}{\sigma_2^2}\right] \tag{35}$$

and:

$$I(f_2, f_1) = \frac{1}{2}\left[\ln\left(\frac{\sigma_1^2}{\sigma_2^2}\right) - 1 + \frac{\sigma_2^2}{\sigma_1^2} + \frac{(\mu_1 - \mu_2)^2}{\sigma_1^2}\right] \tag{36}$$

(Kullback, 1968). Equations 35 and 36 are useful for estimation of $I(f_1, f_2)$ and $I(f_2, f_1)$ if sample estimates of $\mu_1$, $\mu_2$, $\sigma_1$ and $\sigma_2$ are available. In addition, the following properties of the differential relative entropies $I(f_1, f_2)$ and $I(f_2, f_1)$ (for $f_1(x)$ and $f_2(x)$ both Normal) can be deduced from Equations 35 and 36.

1. If $\mu_1 = \mu_2$ and $\sigma_1 = \sigma_2$, $I(f_1, f_2) = I(f_2, f_1) = 0$; both $I(f_1, f_2)$ and $I(f_2, f_1) > 0$ otherwise.

2. For given $\sigma_1$ and $\sigma_2$, both $I(f_1, f_2)$ and $I(f_2, f_1)$ increase as the magnitude of the difference $\mu_1 - \mu_2$ increases (since indicator variables are normally calibrated so that cases tend to have larger indicator scores than controls, normally $\mu_1 > \mu_2$, but in any case the term $\mu_1 - \mu_2$ is squared in Equations 35 and 36 so the sign of $\mu_1 - \mu_2$ is immaterial here).

3. For given $\mu_1$ and $\mu_2$:

    a. if $\sigma_1 = \sigma_2$, $I(f_1, f_2) = I(f_2, f_1)$;

    b. if $\sigma_1 > \sigma_2$, $I(f_1, f_2) > I(f_2, f_1)$;

    c. if $\sigma_1 < \sigma_2$, $I(f_1, f_2) < I(f_2, f_1)$.

To see how 3b and 3c come about, all you need is Love (1980).

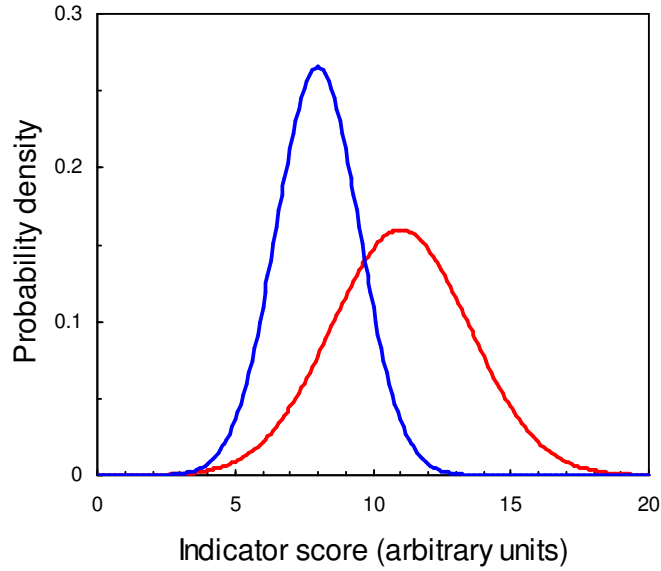

**FIG. 12.** Distributions of indicator scores. In this hypothetical example, the distribution of indicator scores for cases $f_1(x)$ is Normal with mean $\mu_1 = 11$ and standard deviation $\sigma_1 = 2.5$. The distribution of indicator scores for controls $f_2(x)$ is Normal with mean $\mu_2 = 8$ and $\sigma_2 = 1.5$. From Equations 33 and 34, respectively, $I(f_1,f_2) = 2.378$ nits and $I(f_2,f_1) = 0.911$ nits.

The usual way of summarizing distributions of indicator scores as shown in Fig. 12 is to plot an ROC curve. Here we write the cumulative distribution functions $F_1(t) = \int_{-\infty}^{t} f_1(x)\,dx$ and $F_2(t) = \int_{-\infty}^{t} f_2(x)\,dx$, then:

$$TPP = 1 - F_1(t) \tag{37}$$

$$FPP = 1 - F_2(t) \tag{38}$$

and as usual, the ROC curve is a graphical plot of *TPP* against *FPP* as a threshold cut-off $t$ is varied over an appropriate range of indicator scores (Fig. 13). Note, in passing, that graphical plots of binormal indicators have some properties that we do not pursue in the present context (see, for example, Somoza and Mossman, 1991). Here, we can now augment the ROC curve with a plot of relative entropies (Fig. 13). For each pair of (*FPP*,*TPP*) coordinates, a value of $I(D_1,D_2)$ can be calculated from Equation 27, and a value of $I(D_2,D_1)$ from Equation 28 (the discrete forms are calculated because the calculation of *TPP* and *FPP* effectively discretizes the distributions of indicator scores for cases and controls, respectively). As noted previously, the ROC curve is a useful summary of the distributions of indicator scores for cases and controls when these distributions are discretized into $n=2$ categories of prediction outcome, but not for $n>2$.

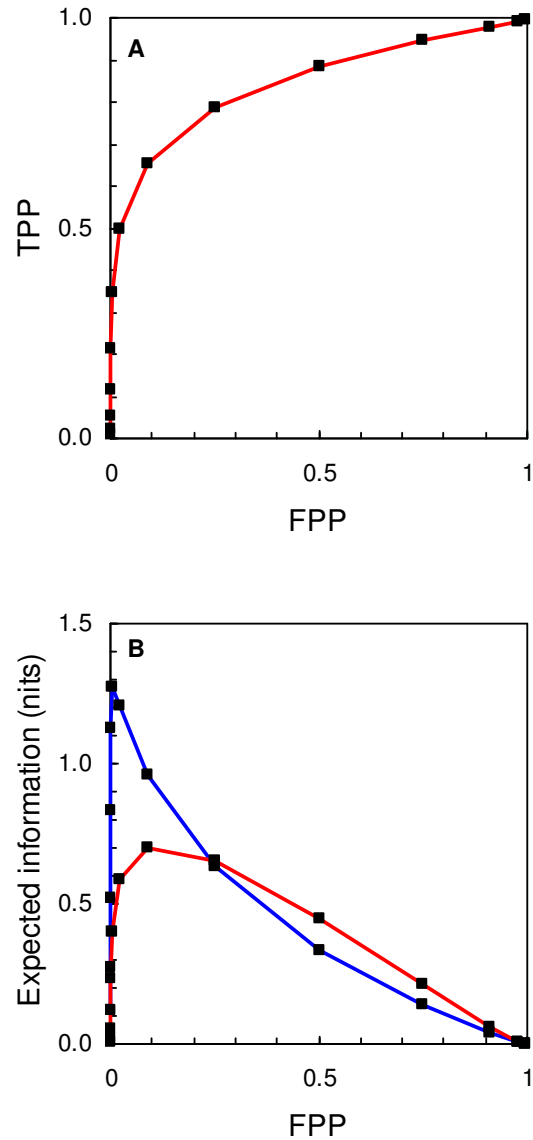

**FIG. 13. A**. The ROC curve derived from the distributions of indicator scores for cases and controls shown in Fig. 12. *TPP* and *FPP* are calculated from Equations 37 and 38, respectively, with $t$ taking integer values 4,5,...,17. **B**. Relative entropies $I(D_1,D_2)$ (blue line) and $I(D_2,D_1)$ (red line) are calculated from Equations 27 and 28, respectively, for each pair of ($FPP$,$TPP$) values used to plot the ROC curve.

### 4.2. The relative distribution

The relative distribution has been discussed in detail by Handcock and Morris (1998, 1999). To begin, we denote the inverse cumulative distribution function:

$$Q(r) = F^{-1}(r)$$

where $r$ is a probability, $0 \le r \le 1$. So, for example, for $f_2(x) \sim N(8,1.5)$ (Fig. 12), $Q_2(0.5) = 8$ is the median of the distribution since:

$$F_2(8) = \int_{-\infty}^{8} f_2(x)\,\mathrm{d}x = 0.5$$

and it follows, by definition, that:

$$F_2(Q_2(r)) = \int_{-\infty}^{Q_2(r)} f_2(x)\,\mathrm{d}x = r .$$

If we call the underlying distribution of indicator scores for controls $f_2(x)$ the reference distribution and the underlying distribution of indicator scores for cases $f_1(x)$ the comparison distribution, the (cumulative) *relative distribution* $G_1(r)$ of cases to controls is then:

$$G_1(r) = F_1(Q_2(r)) = \int_{-\infty}^{Q_2(r)} f_1(x)\,\mathrm{d}x \tag{39}$$

(Fig. 14). The corresponding probability density function $g_1(r)$, called the *relative density*, is:

$$g_1(r) = \frac{f_1(Q_2(r))}{f_2(Q_2(r))} \tag{40}$$

(Fig. 14), and we note $G_1(1) = \int_0^1 g_1(r)\,\mathrm{d}r = 1$. Similarly, if we call underlying distribution of indicator scores for cases $f_1(x)$ the reference distribution and the underlying distribution of indicator scores for controls $f_2(x)$ the comparison distribution, the cumulative relative distribution $G_2(r)$ of controls to cases is:

$$G_2(r) = F_2(Q_1(r)) = \int_{-\infty}^{Q_1(r)} f_2(x)\,\mathrm{d}x \tag{41}$$

(Fig. 14), and then the corresponding probability density function $g_2(r)$ is:

$$g_2(r) = \frac{f_2(Q_1(r))}{f_1(Q_1(r))} \tag{42}$$

(Fig. 14), with $G_2(1) = \int_0^1 g_2(r)\,\mathrm{d}r = 1$.

The relative densities $g_1(r)$ and $g_2(r)$, respectively, are related to the relative entropies $I(f_1,f_2)$ (with cases as the comparison distribution and controls as the reference distribution as in Equation 31 and $I(f_2,f_1)$ (with controls as the comparison distribution and cases as the reference distribution as in Equation 32) as follows:

$$I(f_1, f_2) = \int_0^1 g_1(r)\log(g_1(r))\,\mathrm{d}r \tag{43}$$

$$I(f_2, f_1) = \int_0^1 g_2(r)\log(g_2(r))\,\mathrm{d}r \tag{44}$$

(Mielniczuk, 1992). As previously, the choice of base of logarithm serves only to define the measurement scale.

There is a relationship between the ROC curve and the relative distributions $G_1(r)$ and $G_2(r)$. Bear in mind here that $r$ does not represent a single numerical scale, but is denominated in quantiles of the reference distribution. Then, starting from Equation 39 (so that the distribution of indicator scores for controls is the reference distribution), the graphical plot of *TPP* against *FPP* is most easily produced by plotting $1-G_1(r)$ against $1-r$. Starting from Equation 41 (so that the distribution of indicator scores for cases is the reference distribution), the graphical plot required to reproduce the ROC curve is $1-r$ against $1-G_2(r)$.

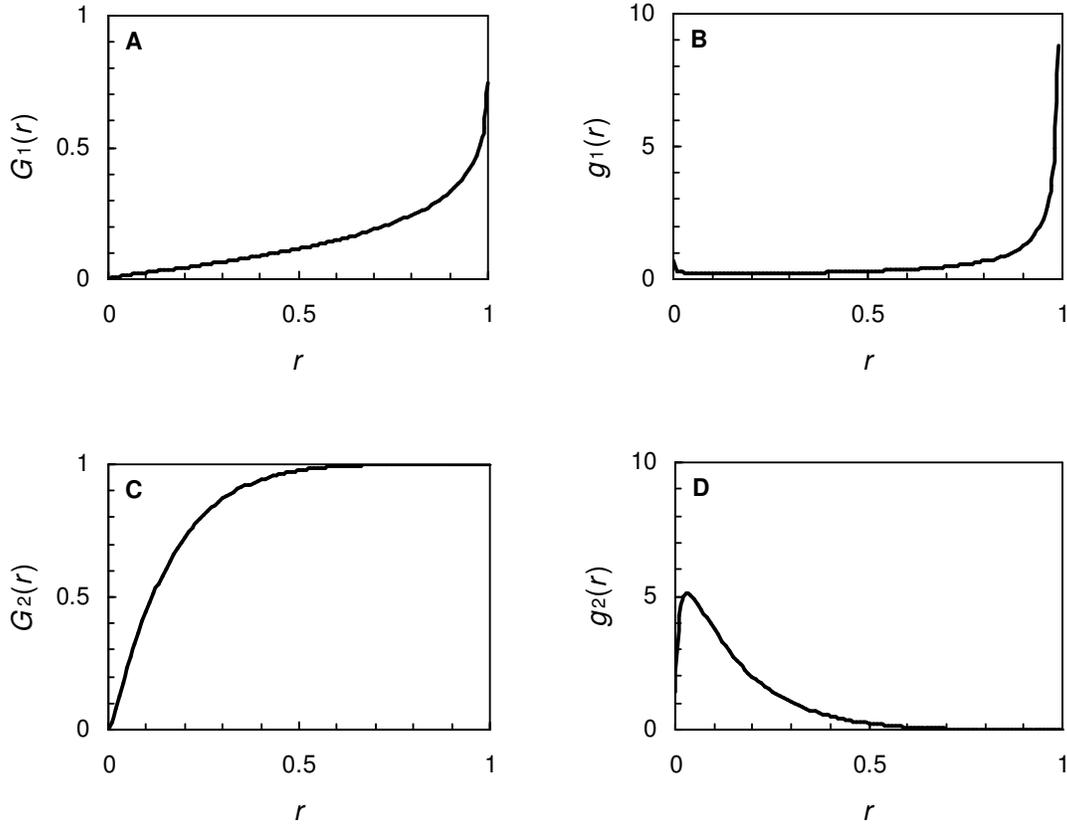

**FIG. 14.** The relative distribution. The distribution of indicator scores for cases $f_1(x)$ and the distribution of indicator scores for controls $f_2(x)$ are as given in Fig. 12. In A and B, $f_2(x)$ is the reference distribution and $f_1(x)$ is the comparison distribution. In C and D, $f_1(x)$ is the reference distribution and $f_2(x)$ is the comparison distribution. **A.** The cumulative relative distribution $G_1(r)$ (Equation 39). **B.** The relative density $g_1(r)$ (Equation 40). **C.** The cumulative relative distribution $G_2(r)$ (Equation 41). **D.** The relative density $g_2(r)$ (Equation 1.42).

For mathematical tractability, the most convenient form relating the ROC curve to the relative distribution (with the distribution of indicator scores for controls as the reference distribution) is a plot of $1 - G_1(1-r)$ as a function of $r$ (see, for example, Li et al., 1996) (Fig. 15). The relative density $g_1(1-r)$ represents the slope of the ROC curve (Fig. 15). When the distribution of indicator scores for cases is the reference distribution, the relative distribution can be related to the ROC curve by a plot of $r$ as a function of $1 - G_2(1-r)$. Now, the relative density $g_2(1-r)$ represents the *reciprocal* of the slope of the ROC curve, and a plot of $1/g_2(1-r)$ as a function of $1 - G_2(1-r)$ represents the slope of the ROC curve (Fig. 15). The relative densities $g_1(1-r)$ and $g_2(1-r)$ retain the previously-mentioned properties of $g_1(r)$ and $g_2(r)$, so:

$$\int_0^1 g_1(1-r)\,\mathrm{d}r = 1,$$

$$\int_0^1 g_1(1-r)\log\bigl(g_1(1-r)\bigr)\,\mathrm{d}r = I(f_1, f_2),$$

$$\int_0^1 g_2(1-r)\,\mathrm{d}r = 1, \text{ and}$$

$$\int_0^1 g_2(1-r)\log\big(g_2(1-r)\big)\,\mathrm{d}r = I\big(f_2,f_1\big).$$

Thus, the relative densities $g_1(1-r)$ and $g_2(1-r)$ capture, in the relative entropies $I(f_1,f_2)$ and $I(f_2,f_1)$, the information characteristics of an indicator as depicted in Fig. 12, and also characterize the slope of the ROC curve derived from that indicator. For the indicator characterized by Fig. 12, the continuous relative entropies $I(f_1,f_2) = 2.378$ nits and $I(f_2,f_1) = 0.991$ nits.

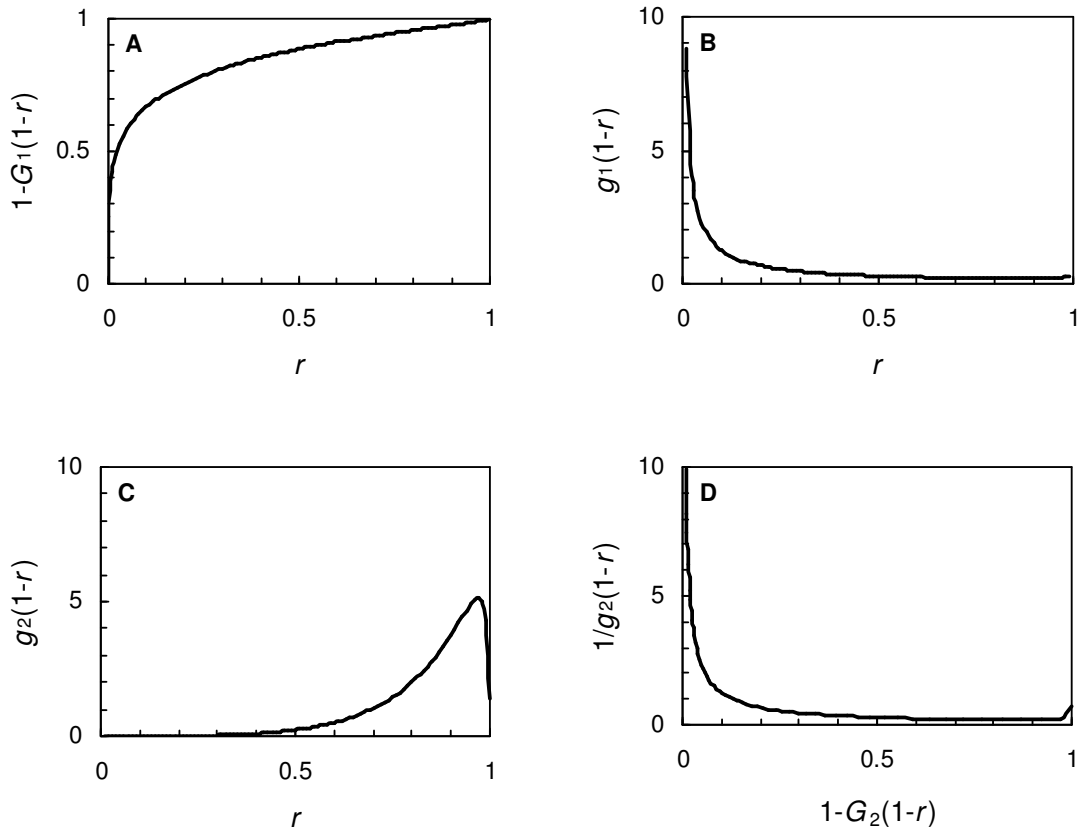

**FIG. 15.** How the relative distribution is related to the ROC curve. The distribution of indicator scores for cases $f_1(x)$ and the distribution of indicator scores for controls $f_2(x)$ are as given in Fig. 12. In A and B, $f_2(x)$ is the reference distribution and $f_1(x)$ is the comparison distribution. In C and D, $f_1(x)$ is the reference distribution and $f_2(x)$ is the comparison distribution. **A.** The cumulative distribution $1-G_1(1-r)$ is the ROC curve (see Fig. 13). **B.** The probability density function $g_1(1-r)$. **C.** The probability density function $g_2(1-r)$. **D.** $1/g_2(1-r)$ as a function of $1-G_2(1-r)$.

## 4.3. A binary predictor based on the relative density $g_1(1-r)$

We are now ready to define a two-outcome-category predictor by specifying a decision rule to operate with our indicator variable (Fig. 12). The indicator variable is characterized by the relative density $g_1(1−r)$ (Fig. 15) and the binary decision rule is embodied in a single threshold value $t$ which is a quantile of the reference distribution (i.e., $f_2(x)$ in this case). For numerical example, we take $t=0.5$, so that the threshold is the median value of the reference distribution, which is $Q_2(0.5)=8$ on the measurement scale of the indicator variable. The predictor is characterized by calculating the average value of $g_1(1−r)$ over the range $0…t$ (Equation 45) and over the range $t…1$ (Equation 46):

$$\frac{1}{t} \cdot \int_0^t g_1(1−r)\,\mathrm{d}r = LR_1 \tag{45}$$

$$\frac{1}{1−t} \cdot \int_t^1 g_1(1−r)\,\mathrm{d}r = LR_2 \tag{46}$$

On the left hand side of Equation 45, the numerator is $TPP$ and the denominator is $FPP$, so the average value of $g_1(1−r)$ over the range $0…t$ is the likelihood ratio $LR_1$ (as in section 1.4). For the indicator characterized by Fig. 12 operated with a threshold value $t=0.5$, $TPP=0.885$, $FPP=0.5$ and $LR_1=1.77$. On the left hand side of Equation 46, the numerator is $FNP$ and the denominator is $TNP$, so the average value of $g_1(1−r)$ over the range $t…1$ is the likelihood ratio $LR_2$ (section 1.4). For the indicator characterized by Fig. 12 operated with a threshold value $t=0.5$, $FNP=0.115$, $TNP=0.5$ and $LR_2=0.23$. Plotted as a probability distribution (Fig. 16), the likelihood ratio function $LR_i$, ($i=1,2$) is a discretization of the continuous relative density function $g_1(1−r)$. The corresponding cumulative distribution function (Fig. 16) is Biggerstaff's (2000) likelihood ratios graph.

For the predictor characterized by Fig. 16, which is the indicator characterized by Fig. 12 operated at a threshold of $FPP=0.5$, the corresponding discrete relative entropies are, respectively, $I(D_1,D_2) = 0.336$ nits and $I(D_2,D_1) = 0.449$ nits (see Fig. 13B). Since in this case $I(D_2,D_1) > I(D_1,D_2)$, this predictor provides, on average, more information for discrimination in favour of $D_2$ (controls) against $D_1$ (cases) than for $D_1$ against $D_2$ (section 3.1). Note, however, that this is an outcome of the particular choice of threshold. For the same indicator, we could choose a threshold that resulted in $I(D_1,D_2) > I(D_2,D_1)$, and therefore a predictor that provides more information for discrimination in favour of $D_1$ (cases) against $D_2$ (controls) than for $D_2$ against $D_1$. For the indicator characterized by Fig. 12, $I(D_1,D_2) > I(D_2,D_1)$ when $FPP$ is below about 0.25, and $I(D_2,D_1) > I(D_1,D_2)$ above this value (Fig. 13B).

In the approach outlined here, a predictor has been characterized by a likelihood ratio function obtained by discretizing the continuous relative density function $g_1(1−r)$. This relative density function is a characteristic of the indicator (as represented by Fig. 12 in the present example).

### 4.4. A predictor with multiple outcome categories based on the relative density $g_1(1−r)$

We now define an $n$-outcome-category predictor ($n≥2$) by specifying a decision rule incorporating $n$-1 thresholds to operate with our indicator variable (Fig. 12). We can take $n≥2$ (not just $n>2$) because the two-outcome-category case is covered by the general treatment here, although we have already seen that $n=2$ can be treated as a special case (section 4.3). Now, the indicator variable is characterized by the relative density $g_1(1−r)$ (Fig. 15) and the decision rule is embodied in threshold values $t_1,… t_i, … t_{n-1}$ which are quantiles of the reference distribution (i.e., $f_2(x)$ in this case).

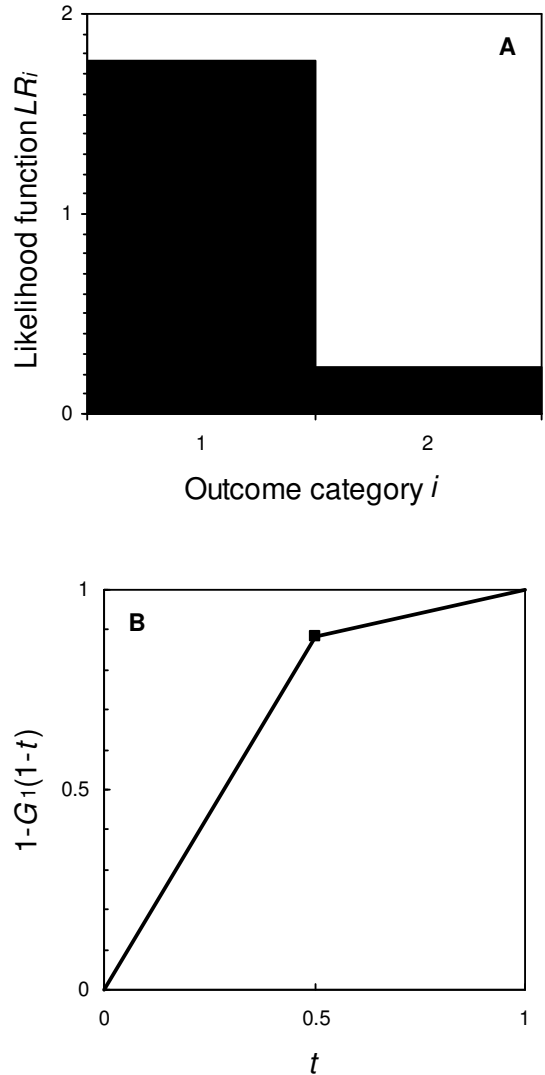

**FIG. 1.16.** Likelihood ratios. **A.** The likelihood ratio function $LR_i$, ($i$=1,2) is a discretization of the continuous relative density function $g_1(1-r)$ derived from the indicator characterized by Fig. 12, with a decision rule embodied in a single threshold value of $r$, denoted $t$=0.5, which is the median of the reference distribution $f_2(x)$. **B.** The corresponding cumulative distribution function.

For numerical example, we take $n$=4, $t_1$=0.25, $t_2$=0.5, $t_3$=0.75, so that the thresholds are the first quartile, the median (i.e., the second quartile), and the third quartile of the reference distribution, respectively. On the measurement scale of the indicator variable, the threshold values are given by $Q_2(t_i)$, $i$=1…$n$-1. This predictor is characterized by calculating average values of $g_1(1-r)$ as follows:

$$\frac{1}{t_1} \cdot \int_0^{t_1} g_1(1-r)\, \mathrm{d}r = LR_1,$$

$$\frac{1}{t_2 - t_1} \cdot \int_{t_1}^{t_2} g_1(1-r)\, \mathrm{d}r = LR_2,$$

$$\frac{1}{t_3 - t_2} \cdot \int_{t_2}^{t_3} g_1(1-r)\,dr = LR_3,$$

$$\frac{1}{1 - t_3} \cdot \int_{t_3}^{1} g_1(1-r)\,dr = LR_4.$$

For this example, $LR_1 = 3.147$, $LR_2 = 0.393$, $LR_3 = 0.243$ and $LR_4 = 0.217$. Plotted as a probability distribution (Fig. 17), the likelihood ratio function $LR_i$, ($i=1,\ldots,4$) is a discretization of the continuous relative density function $g_1(1-r)$. The corresponding cumulative distribution function (Fig. 17) is a generalization of Biggerstaff's (2000) likelihood ratios graph.

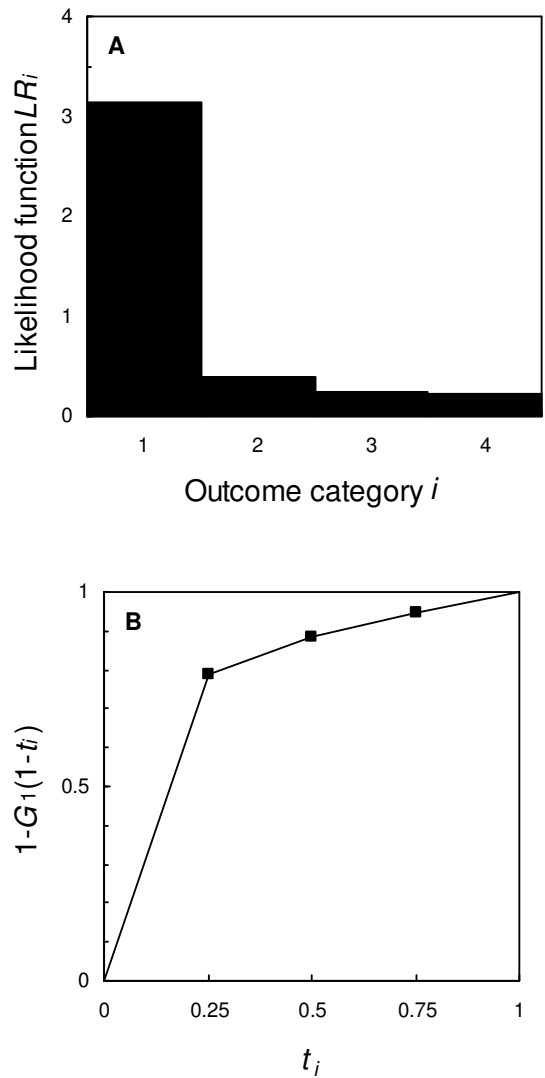

**FIG. 1.17.** Likelihood ratios. **A**. The likelihood ratio function $LR_i$, ($i=1,\ldots,n$, $n=4$) is a discretization of the continuous relative density function $g_1(1-r)$ derived from the indicator characterized by Fig. 1.12, with a decision rule embodied in $n-1$ threshold values of $r$, denoted $t_i$ ($i=1,\ldots,n-1$), which in this example are respectively the 1[st], 2[nd] and 3[rd] quartiles of the reference distribution $f_2(x)$. **B**. The corresponding cumulative distribution function.

Recall that for the two-outcome-category predictor characterized by Fig. 16, which is the indicator characterized by Fig. 12 operated with a single threshold at the median of the reference distribution (section 4.3), the discrete relative entropies are $I(D_1,D_2) = 0.336$ nits and $I(D_2,D_1) = 0.449$ nits. Now, for the four-outcome-category predictor characterized by Fig. 17, which is the indicator characterized by Fig. 12 operated with three thresholds, at the quartiles of the reference distribution, the discrete relative entropies are $I(D_1,D_2) = 0.642$ nits and $I(D_2,D_1) = 0.683$ nits. Thus it is still the case that $I(D_2,D_1) > I(D_1,D_2)$, but both relative entropies are larger than for the two-outcome-category predictor characterized by Fig. 16.

As in section 4.3, a predictor has been characterized by a likelihood ratio function obtained by discretizing the continuous relative density function $g_1(1-r)$. This relative density function is a characteristic of the indicator (as represented by Fig. 12 in the present example). The function $g_1(1-r)$ provides a basis for obtaining the interval likelihood ratios required for a multiple-outcome-category predictor. This is useful because methods based on the ROC curve can only be used to obtain likelihood ratios when discretization of the distributions of indicator scores for cases and controls results in two categories of prediction outcomes, as in Table 7. Discretization of $g_1(1-r)$ provides a basis for characterizing a predictor with *two or more* categories of prediction outcomes.

### 4.5.  A summary of some useful properties of the relative density

- The continuous relative density functions $g_1(1-r)$ and $g_2(1-r)$ $(0 \leq r \leq 1)$ characterize a continuous indicator variable for which there are descriptions of the separate distributions of indicator scores for cases and for controls.

- Both $g_1(1-r)$ and $g_2(1-r)$ are probability density functions.

- The differential relative entropies $I(f_1,f_2)$ and $I(f_2,f_1)$ (Equations 43 and 44, respectively) characterize the information properties of an indicator for which we have $g_1(1-r)$ and $g_2(1-r)$.

- The ROC curve is a familiar property of an indicator variable. The relative density $g_1(1-r)$ represents the slope of the ROC curve. The ROC curve can be obtained from the cumulative relative distribution $G_1(1-r)$ most conveniently by a plot of $1-G_1(1-r)$ as a function of $r$.

- A widely-used single figure summary of the properties of an indicator variable is the area under the ROC curve (see Hanley and McNeil (1982)). However, it can be advantageous to have a summary in the form $I(f_1,f_2)$ and $I(f_2,f_1)$ as not all indicators are symmetrical in terms of their properties relating to cases and to controls. In other words, ROC curves with the same area under the curve do not necessarily characterize identical indicators.

- To obtain a binary predictor, we discretize $g_1(1-r)$ with single threshold. The resulting predictor is characterized by two likelihood ratios (Equations 45 and 46). Together these represent a discrete likelihood function that is a probability mass function. We can characterize the information properties of the predictor by calculating the discrete relative entropies $I(D_1,D_2)$ and $I(D_2,D_1)$ (Equations 23 and 24). These values are comparable with the differential relative entropies $I(f_1,f_2)$ and $I(f_2,f_1)$ for the indicator, which we can think of as the limiting forms.

- The cumulative version of the discrete likelihood function for a binary predictor is the likelihood ratios graph described by Biggerstaff (2000).

- In general, if we discretize $g_1(1-r)$ with $n-1$ thresholds ($n \geq 2$), we obtain a predictor with $n$ outcome categories. This predictor is characterized by $n$ interval likelihood ratios, and together these represent a discrete likelihood function that is a probability mass function. As is the case for a binary predictor, we can characterize the information properties of the $n$-outcome-category predictor by calculating the discrete relative entropies $I(D_1,D_2)$ and $I(D_2,D_1)$ (Equations 23 and 24). These values are comparable with the differential relative entropies $I(f_1,f_2)$ and $I(f_2,f_1)$ for the indicator, which we can think of as the limiting forms. Note that interval likelihood ratios as calculated in section 3.2 are ratios of two probability mass functions, but in that case the resulting discrete likelihood function is not, in general, a itself probability mass function.

- The cumulative version of the discrete likelihood function for a predictor with $n$ outcome categories ($n \geq 2$) is a generalization of the likelihood ratios graph described by Biggerstaff (2000) for the case of $n=2$ .

- The literature on interval likelihood ratios often resorts to a contrived derivation of interval likelihoods from an ROC curve (see, for example, Black and Armstrong (1986), Choi (1998), Sonis (1999) and Brown and Reeves (2003)). To be clear: an ROC curve is a property of an indicator variable, not of a predictor. A predictor (i.e., a diagnostic test) comprises both an indicator variable *and* a decision rule that defines the operational threshold(s) for the categories of prediction outcome. If a decision rule is binary, its two likelihood ratios may, if required for convenience, be written in terms of sensitivity and specificity. If a decision rule has more than two categories of prediction outcome, the interval likelihood ratios are written and calculated as in Equation 31. In this case, sensitivity and specificity are not defined, notwithstanding the literature cited above. Now, however, we have a valid method by which to calculate the interval likelihood ratios directly from the relative density, without reference to sensitivity and specificity. If we discretize $g_1(1-r)$ with $n-1$ thresholds, we obtain a predictor with $n$ outcome categories ($n \geq 2$), characterized by $n$ interval likelihood ratios.